\documentclass[11pt,a4paper]{article}
\pdfoutput=1
\usepackage{jcappub}
\usepackage[T1]{fontenc}    
\usepackage{graphicx}
\usepackage{slashed}
\usepackage[compat=1.1.0]{tikz-feynman}
\usepackage{placeins}
\usepackage[utf8]{inputenc} 

\newcommand{\be}{\begin{equation}}
\newcommand{\ee}{\end{equation}}
\newcommand{\beqa}{\begin{eqnarray}}
\newcommand{\eeqa}{\end{eqnarray}}

\newcommand{\M}{\text{{\rm M}\o l}}

\def\Slashnew#1{#1\kern-0.55em\raise.05ex\hbox{/}}
\def\slashnew#1{#1\kern-0.5em\raise.05ex\hbox{{$\scriptstyle /$}}}
\def\lsim{\mathrel{\raise.3ex\hbox{$<$\kern-.75em\lower1ex\hbox{$\sim$}}}}
\def\gsim{\mathrel{\raise.3ex\hbox{$>$\kern-.75em\lower1ex\hbox{$\sim$}}}}

\def\sfrac#1#2{\ensuremath{\textstyle \frac{#1}{#2}}}

\def\emph#1{{\em #1}}

\def\ie{{\em i.e.~}}

\def\S{{\scriptscriptstyle \rm S}}

\def\h{{\scriptscriptstyle \rm h}}
\def\f{{\scriptscriptstyle \rm f}}

\def\dy{{\bf y}}
\def\DM{{\scriptscriptstyle \rm DM}}
\def\mDM{{m_\DM}}
\def\MPl{{M_{\scriptscriptstyle \rm Pl}}}

\title{Precision calculations of dark matter relic abundance}

\author[a]{Kalle Ala-Mattinen}
\author[b,c,d]{Kimmo Kainulainen}
\affiliation[a]{Department of Physics, P.O.Box 64, FI-00014 University of Helsinki, Finland}
\affiliation[b]{Department of Physics, PL 35 (YFL), 40014 University of Jyv\"askyl\"a, Finland}
\affiliation[c]{Helsinki Institute of Physics, PL 64, 00014 University of Helsinki, Finland}
\affiliation[d]{Theoretical Physics Department, CERN, 1211 Geneva 23, Switzerland}

\emailAdd{kalle.ala-mattinen@helsinki.fi}
\emailAdd{kimmo.kainulainen@jyu.fi} 

\abstract{The dark matter annihilation channels sometimes involve sharp resonances. In such cases the usual momentum averaged approximations for computing the DM abundance may not be accurate. We develop an easily accessible momentum dependent framework for computing the DM abundance accurately and efficiently near such features. We apply the method to the case of a singlet scalar dark matter $s$ interacting with SM through higgs portal $\lambda_{hs}s^2 h^2$ and compare the results with different momentum averaged methods. The accuracy of the latter depend strongly on the strength of the elastic interactions and corrections are large if WIMP has negligible interactions beyond the main annihilation channel. In the singlet scalar model however, the standard model scatterings induce an efficient kinetic equilibrium that validates the momentum averaged computation to 20 per cent accuracy. We update the current extent of the allowed region in the light singlet scalar dark matter to $m_\S \in [56,62.5]$ GeV.}


\subheader{CERN-TH-2019-214}

\begin{document}
\maketitle

%
%

%
\section{Introduction}
%

The nature of the dark matter (DM) in the universe remains an unsolved mystery. The most popular candidate for DM is some weakly interacting massive particle. Recently there has been a lot of interest in a class of models where the dark sector interacts with the standard model (SM) through a higgs portal~\cite{McDonald:1993ex,Cline:2013gha}. Generic to these models is that the DM abundance can be adjusted correctly, avoiding all experimental constraints, just below the higgs pole. However, because the SM higgs is a very sharp resonance, computing the DM abundance near its pole, \ie when $2\mDM \lsim m_H$, is more involved than is perhaps usually appreciated. Of course the higgs resonance may not be the only one relevant for the DM production. Other resonances associated with the $Z$-boson or new exotic gauge bosons or new scalars are frequently encountered in the model landscape. In all these cases computing DM abundance requires extra care and the results obtained here can be applied. 

What makes narrow resonances challenging for momentum averaged methods is that the implicit assumption they make, of elastic scatterings being fast enough to keep the system in kinetic equilibrium, may not hold. Annihilation processes can then lead to a significant distortion of the phase space distributions, reducing the number of momentum configurations consistent with the resonance. When this happens, momentum averaged methods, that assume kinetic equilibrium, can lead to an overestimate of the annihilation rate and an underestimation of the DM density. 

We start by setting up the generic Boltzmann equations for the dark matter annihilation problem in section~\ref{sec:mom-ave}. We then review the derivation of the momentum averaged Zel$^\prime$dovich-Okun-Pikelner-Lee-Weinberg~\cite{Zeldovich:1965,Lee:1977ua} (ZOPLW) equation and several approximation schemes to solve it. We then apply these methods to the singlet scalar DM coupled to the Standard Model via higgs portal (the SSM model). We compute the singlet DM abundance and discuss the range of validity of different approximations. We stress that the semi-analytical solution developed in ref.~\cite{Cline:2013gha}, is always within ${\cal O}(1\%)$ agreement with the full numerical solution of ZOPLW equations.

In section~\ref{sec:MBapproximation} we develop a momentum dependent scheme to solve the DM abundance accurately and efficiently. The novel part of the method is the implementation of a generalised relaxation approximation for the back-reaction terms in the elastic collision integrals. Back-reaction terms are multidimensional integrals whose direct evaluation is not practical. In our method all collision terms are reduced to generic one-dimensional integrals over the relevant CM-frame cross sections, which can be evaluated and fitted before the numerical integration of the partial differential equations. Our final equations take form of a set of coupled ZOPLW equations for the discretised momentum modes and for an arbitrary number of interacting species. These equations are one of the main results of this paper. They should be useful also in other applications with non-equilibrium dynamics, such as scenarios with non-thermal DM or particle wall interactions during electroweak phase transition.

In section~\ref{sec:results_wimp} we carefully analyse the DM abundance of a thermal DM near the resonance in the SSM model. We show that without elastic interactions the momentum dependent code can give up to an order of magnitude larger DM abundance than does the best momentum averaged method. When elastic interactions are included however, the momentum dependent calculation gets very close to the momentum averaged one; the residual difference in the DM abundance is typically 20-30 per cent. The self scatterings play no essential role in reaching the kinetic equilibrium in the SSM; it is mainly established by the elastic scatterings with SM particles. We update the current extent of the allowed region of the light DM in the SSM to be $m_\S \in [56,62.5]$ GeV. We also show that a DM throughout this range can be discovered in a direct detection experiment whose sensitivity only slightly exceeds the neutrino floor. 

In section~\ref{sec:results_fimp}, we consider the feebly interacting dark matter (FIMP) limit. We again consider the SSM model and compare the momentum averaged and the momentum dependent methods. We find that while FIMPs are never in thermal equilibrium, they are produced at all times near kinetic equilibrium and the momentum averaged method is again accurate at 20 per cent level.  Finally, in section~\ref{sec:conclusions}, we give our conclusions and outlook.

\section{The Boltzmann equation}
\label{sec:mom-ave}
%

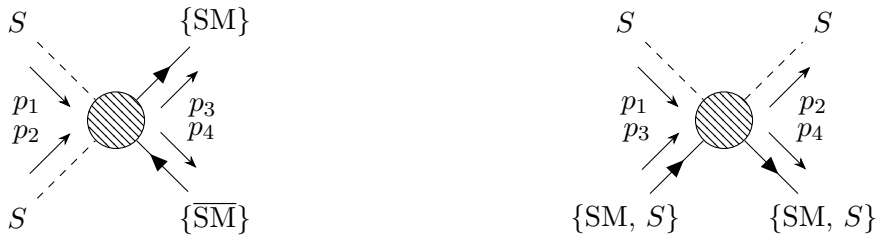
\begin{figure}[t]
\begin{tikzpicture}
  \def\leglength{1.3}
  \hspace{1cm}
  \begin{feynman}
    \vertex[blob] (m)  {};
    \vertex (a) at (-\leglength, \leglength) {$S$};
    \vertex (b) at (-\leglength,-\leglength) {$S$};
    \vertex (c) at ( \leglength, \leglength) {\{${\rm SM}$\}};      
    \vertex (d) at ( \leglength,-\leglength) {\{$\overline{\rm SM}$\}};
    \diagram* {
      (a) -- [scalar, momentum'  = \(p_1\)] (m) -- [fermion, 	  momentum' = \(p_3\)] (c),
      (b) -- [scalar, momentum = \(p_2\)] (m) -- [anti fermion, momentum = \(p_4\)] (d),
    };
  \end{feynman}
  \hspace{8cm}
    \begin{feynman}
    \vertex[blob] (m)  {};
    \vertex (a) at (-\leglength, \leglength) {$S$};
    \vertex (b) at (-\leglength,-\leglength) {\{${\rm SM}$, $S$\}};
    \vertex (c) at ( \leglength, \leglength) {$S$};      
    \vertex (d) at ( \leglength,-\leglength) {\{${\rm SM}$, $S$\}};
    \diagram* {
      (a) -- [scalar, momentum'  = \(p_1\)] (m) -- [scalar, momentum' = \(p_2\)] (c),
      (b) -- [fermion, momentum = \(p_3\)] (m) -- [fermion, momentum = \(p_4\)] (d),
    };
  \end{feynman}
\end{tikzpicture}
\caption{Inelastic and elastic $2 - 2$ collision processes. Standard Model contributions are collectively referred as ${\rm SM}$ and the relevant contributions to our analysis are: $\tau , c , t , b , Z , W , h$ in the inelastic channel and $\mu,\tau,s,c,b$ in the elastic channel. Section~\ref{sec:MBapproximation} follows the indexing conventions of these diagrams.}
\label{fig:scatter}
\end{figure}

The Boltzmann equation for the scalar distribution function $f(p_1,t)$ in the flat Friedmann-Robertson-Walker spacetime is
\begin{equation}
(\partial_t - p_1H\partial_{p_1})f(p_1,t) = \hat C_{\rm E}(p_1,t) + \hat C_{\rm I}(p_1,t),
\label{eq:BoltzmannEq}
\end{equation}
where $\hat C_{\rm E}(p_1,t)$ and $\hat C_{\rm I}(p_1,t)$ are the elastic and inelastic collision integrals respectively.  Elastic collisions are responsible for maintaining the kinetic equilibrium and inelastic collisions the chemical equilibrium. Inelastic collision integral is generically given by
\begin{eqnarray}
\hat C_{\rm I}(p_1,t) &=& \frac{1}{2E_1} \sum_n \int \frac{{\rm d}^3p_2}{(2\pi)^32E_2}  \left[ \prod_{\{i_n\}}
\frac{{\rm d}^3 p_{i_n}}{(2\pi )^32E_{i_n}} \right] (2\pi )^4\delta^4(p_1+p_2- \sum_{\{i_n\}} p_{i_n}) \times
\nonumber \\
&\times & \Big(  \mid {\cal M}_{{\{i_n\}} \rightarrow 12}^{(n)}\mid^2
 \prod_{\{i_n\}} [ f_{i_n}(p_{i_n},t) ]  \; [1+f(p_1,t)] [1+f(p_2,t)] 
\nonumber \\ 
&&- \hskip-0.5mm \mid {\cal M}_{12\rightarrow {\{i_n\}}}^{(n)}\mid^2
                 f(p_1,t)f(p_2,t)\prod_{\{i_n\}}[1+s_{i_n}f_{i_n}(p_{i_n},t)] \; \Big) \,,
\end{eqnarray}
and the elastic one, assuming it is dominated by $2-2$-scattering processes, by
\begin{eqnarray}
\hat C_{\rm E}(p_1,t) &=& \frac{1}{2E_1} \sum_m \int \left[ \prod_{i=2}^4
\frac{{\rm d}^3 p_i}{(2\pi )^32E_{mi}}\right] (2\pi )^4\delta^4(p_1+p_3-p_2-p_4) \times
\nonumber \\
&\times & \left( \;\; \mid {\cal M}_{24\rightarrow 13}^{(m)}\mid^2
                 f(p_2,t) f_{m4}(p_4,t)[1+f(p_1,t)][1+s_{m3}f_{m3}(p_3,t)] \right. 
\nonumber \\ 
&&-\left. \; \mid {\cal M}_{13\rightarrow 24}^{(m)}\mid^2
                 f(p_1,t)f_{m3}(p_3,t)[1+f(p_2,t)][1+s_{m4}f_{m4}(p_4,t)] \; \right) \,,
\end{eqnarray}
where $E_i = (p_i^2 + m_i^2)^{1/2}$, the indices $n$ and $m$ run through all relevant interaction channels, $f_i(p,t)$ are the momentum- and time-dependent distribution functions of the particle species in question, and $s_n = 1$ (-1) for bosons (fermions). The matrix elements $|{\cal M}_{ij\rightarrow kl}|^2$ are process dependent functions that only depend on the Mandelstam variables $s$, $t$ and $u$. The inelastic and elastic channels are shown schematically in figure~\ref{fig:scatter}. Including additional decay-channels or processes with more than two particles in the final state would be straightforward.

%
\subsection{Momentum integrated equation}
\label{sec:momentum_averaged}
%

\begin{figure}[hbt!]
\begin{center}
\includegraphics[width=0.5\textwidth]{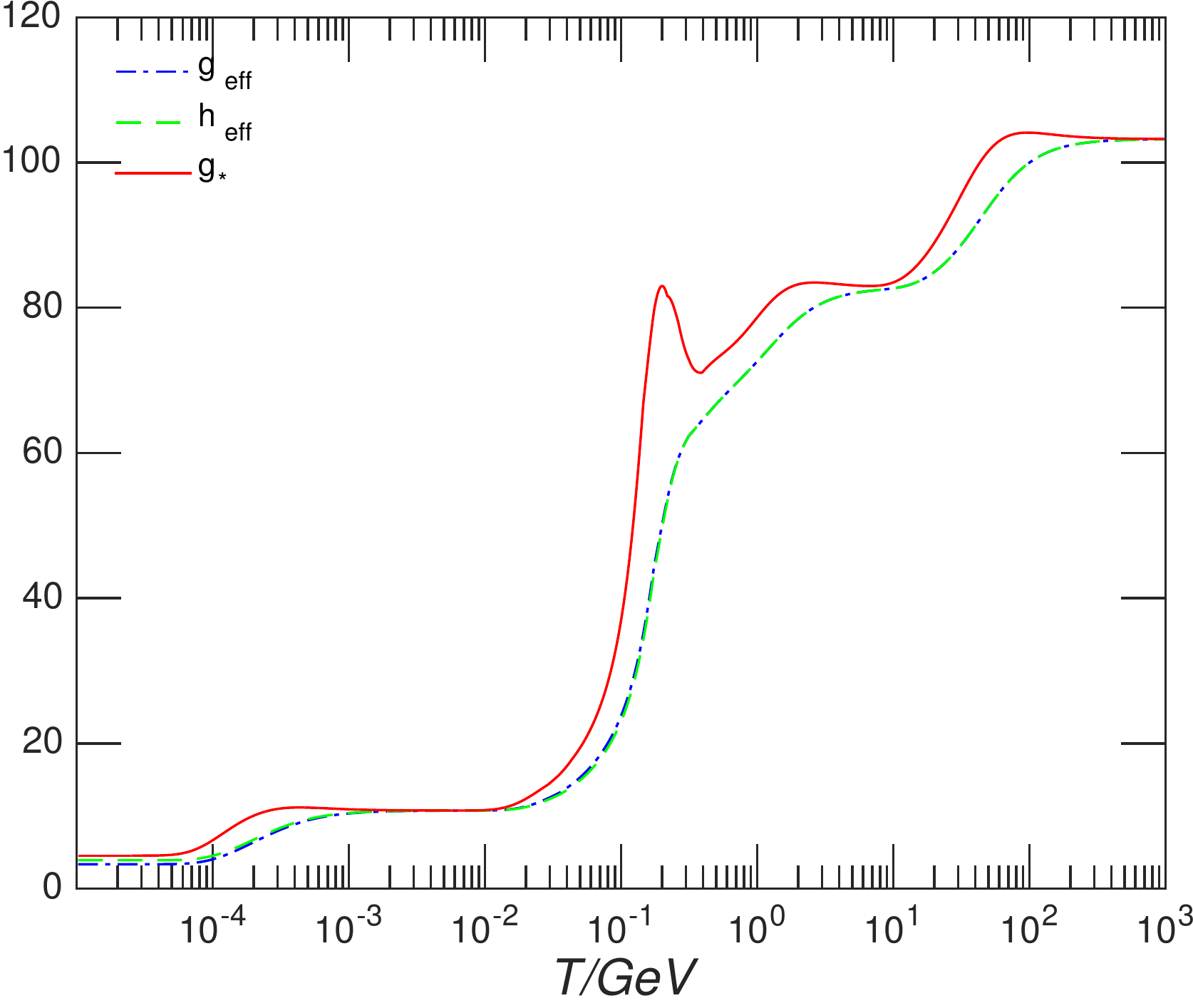} 
\caption{The effective degrees of freedom functions: $g_{\rm eff}$ (dash-dotted), $h_{\rm eff}$ (dashed) and $g_*$ (solid). The feature around $T \approx 175$ MeV is due to latent heat release in the QCD phase transition, which is here modelled by $T_{\rm QCD}=175$ MeV and a linear interpolation between the hadronic phase and the quark phase results over a width $\Delta T_{\rm QCD} = 70$ MeV.}
\label{fig:effdofs}
\end{center}
\end{figure}

A standard approximation in relic density calculations is that particles are in kinetic equilibrium at all times and and that they follow the Maxwell-Boltzmann statistics:
\begin{equation}
f_i(p_i,t) \rightarrow g_i(T)e^{-E_{p_i}/T} \,,
\label{eq:pseudoeqdist}
\end{equation}
with $g_{i,\rm eq}(T)=1$. With these assumptions we can integrate the Boltzmann equation (\ref{eq:BoltzmannEq}) over the initial three momentum ${\bf p}_1$. The elastic collision term $\hat C_{\rm E}$ now vanishes and the momentum-dependent equation reduces to the Zel$^\prime$dovich-Okun-Pikelner-Lee-Weinberg equation~\cite{Zeldovich:1965,Lee:1977ua} for the number density:
\begin{equation}
\partial_t n + 3H n = \langle v_\M \sigma_{\rm I} \rangle (n_{\rm eq}^2 - n^2)\,,
\label{eq:LWEquation}
\end{equation}
where the averaged cross section is the Maxwell-Boltzmann average over the annihilation cross section multiplied by the M\o ller velocity: $v_\M \equiv ((p_1 p_2)^2 - m_1^2m_2^2)^{1/2}/E_1E_2$. It is a simple matter to reduce this quantity to a one dimensional integral over the cross section~\cite{Gondolo:1990dk}:

\begin{equation}
\langle v_\M \sigma_{\rm I} \rangle  = 
\frac{1}{8m_\S^4TK_2(m_\S/T)^2} \int_{4m_\S^2}^\infty {\rm d}s \, \sqrt{s} (s-4m_\S^2) K_1\Big(\frac{\sqrt{s}}{T}\Big)\sigma_{\rm I} (s) \,,
\label{eq:Mollerintegral}
\end{equation}
where $K_i(x)$ are the modified Bessel functions of the second kind and $\sigma_{\rm I}=\sum_n \sigma_{\rm I,n}$, with $n$ labelling separate inelastic processes. Now, it is usual to assume that the universe is expanding adiabatically: $\dot s/s = -3H$, where $s$ is the entropy density. In this case Eq.~(\ref{eq:LWEquation}) can be written as
\begin{equation}
\partial_x Y = Z(x) (Y_{\rm eq}^2 - Y^2)\,,
\label{eq:LWwithY}
\end{equation}
where we defined $x\equiv m/T$ and $Y \equiv n/s$, so that 
\begin{equation}
Y_{\rm eq}(x) = \frac{45}{4\pi^4} \frac{x^2}{h_{\rm eff}(x)}K_2(x) \,,
\label{eq:Yeq}
\end{equation}
and finally
\begin{equation}
Z(x) \equiv \sqrt{\frac{\pi}{45}} g^{1/2}_* \, \frac{m\MPl}{x^2}
\langle v_\M \sigma_{\rm I} \rangle \,.
\label{eq:LWZ}
\end{equation}
Here $\MPl$ is the Planck mass and the function 
\begin{equation}
g^{1/2}_*(T) \equiv \frac{h_{\rm eff}}{\sqrt{g_{\rm eff}}}
\left(1 +\frac{T}{3h_{\rm eff}} \frac{{\rm d} h_{\rm eff}}{{\rm d}T} \right)
\label{eq:gstar}
\end{equation}
depends on the number of effective energy and entropy degrees of freedom defined by: $\rho(T) \equiv \frac{\pi^2}{30}g_{\rm eff}T^4$ and $s(T) \equiv \frac{2\pi^2}{45}h_{\rm eff}T^3$. In the limit of no entropy production and sufficiently high temperatures $g_* \approx g_{\rm eff}$. However, as was stressed already by~\cite{Gondolo:1990dk}, for high-accuracy calculations one should keep the full $g_*$, as the functions do differ in particular near the QCD phase transition. Early work on the number of degrees of freedom functions include~\cite{Srednicki:1988ce} and a careful recent analysis of the effect of QCD transition can be found in~\cite{Laine:2006cp,Drees:2015exa}. We show the functions we are using here in figure~\ref{fig:effdofs}.

%

%
%

%
\subsection{Analytical approximations}
\label{sec:analytical_approx}
%

It is a simple matter to integrate equation (\ref{eq:LWwithY}) numerically. A generic behaviour for $Y$ is that first it follows closely the equilibrium distribution and then abruptly freezes out, typically when the WIMP is non-relativistic: $x_{\rm f} \approx 20-30$. In such case, the ZOPLW equation can be solved analytically to a very high accuracy~\cite{Lee:1977ua,Gondolo:1990dk,Steigman:2012nb,Enqvist:1988we,Cline:2013gha}. In~\cite{Cline:2013gha} it was shown that the solution
\begin{equation}
Y_{\rm today} = \frac{Y_{\rm f}}{1 + Y_{\rm f} A_{\rm f}} \,, 
\quad {\rm where} \quad 
A_{\rm f} = \int_{x_{\rm f}}^\infty {\rm d}x Z(x) \,,
\label{eq:Ytoday}
\end{equation}
and $Y_{\rm f} = (1+\delta_{\rm f})Y_{\rm eq}(x_{\rm f})$, and the freeze-out temperature is solved from
\begin{equation}
x_{\rm f} = \log \left( \frac{ \delta_{\rm f}(2+\delta_{\rm f} ) }{ 1+\delta_{\rm f} } 
\frac{ Z \hat Y_{\rm eq}^2 }{\hat Y_{\rm eq} - \frac{{\rm d}\hat Y_{\rm eq}}{{\rm d}x}} \right)_{x_{\rm f}} \,, \quad {\rm with} \quad \hat Y_{\rm eq} \equiv  e^xY_{\rm eq} \,,
\label{eq:xfout}
\end{equation}
is accurate typically to better than one percent for $\delta_f \approx 1$. Given $Y_{\rm today}$, one can easily find the final abundance:
\begin{eqnarray}
\Omega_{\rm DM} h^2 \simeq 2.7 \times 10^8 \left( \frac{m_i}{\rm GeV} \right) Y_{\rm today} \,.
\label{ecosmo3}
\end{eqnarray}
If $Z(x)$ is only weakly dependent on $x$ one can further approximate $A_{\rm f} \approx x_{\rm f} Z_{\rm f}$. Moreover, one typically finds that $Y_{\rm f}A_{\rm f}\gg 1$, so that $Y_{\rm today} \approx 1/x_{\rm f}Z_{\rm f}$.  Even this approximation is typically accurate to a few per cent. Moreover, when applying these formulae one finds that for typical DM masses $m \approx 10-1000$ GeV the cross section giving the correct relic abundance is almost a constant: $\langle{v_\M\sigma_{\rm I}}\rangle \approx 2.2 \times 10^{-26}\; \rm cm^3/s$~\cite{Steigman:2012nb}. 

To solve the differential equation beyond the analytic approximation, one often uses simplifying approximations for the thermally averaged cross section. Indeed, if $v_\M \sigma_{\rm I}$ approaches a constant in the non-relativistic limit, one may use the threshold approximation:
\begin{equation}
\langle v_\M \sigma_{\rm I} \rangle \approx v_\M^{\rm CM} \sigma_{\rm I}|_{s=4m_\S^2} \equiv (v_\M\sigma_{\rm I})_{\rm th}.
\label{eq:thresholdapp}
\end{equation}
where $v_\M^{\rm CM} = 2\sqrt{1-4m^2/s}$. Threshold approximation often works rather well, but it obviously fails when $v_\M^{\rm CM}\sigma_I$ vanishes at threshold and we shall see that it also fails near sharp resonances. An example of the former is the annihilation of Majorana fermions while a singlet scalar DM near higgs pole is an example of the latter. In contrast, the approximation (\ref{eq:Ytoday}-\ref{eq:xfout}) is {\em always} good one for the ZOPLW equation, independent of how one computes $\langle{v_\M\sigma_{\rm I}}\rangle$,  as long as particles remain in kinetic equilibrium and are non-relativistic at freeze-out.

%
\subsection{Example: singlet scalar DM near higgs pole}
\label{sec:example_m-ave}
%

To be specific, we consider a model with a new scalar singlet field with $Z_2$-symmetry, that  couples to the standard model particles only through the higgs portal:
\begin{equation}
  {\cal L} = \frac12 (\partial_\mu S)^2 - \frac12 \mu_\S^2 S^2 -\frac14 \lambda_\S S^4 - \frac12\lambda_{\rm hs} S^2|H|^2  + {\cal L}_{\rm SM}\;.
\label{spot}
\end{equation}
After electroweak symmetry breaking, the $S$ boson mass receives a mass term $m_\S^2 = \mu_\S^2 + \sfrac12\lambda_{hs} v^2$, where $v=246.2$\,GeV. This model can provide DM over a wide range of parameters. In particular there is an interesting allowed region below the higgs pole~\cite{Cline:2013gha}, the extent of which we now update to $m_\S \in [56,62.5]$ GeV. The precise extent of this allowed region is sensitive to how one computes the relic density. We demonstrate this by solving the ZOPLW equation~(\ref{eq:LWEquation}) both exactly and in the threshold approximation described in previous section. All annihilation cross sections for $s$-boson can be found in  ref.~\cite{Cline:2012hg,Cline:2013gha}\footnote{The $v_{\rm rel}$ defined in  ref.~\cite{Cline:2012hg,Cline:2013gha} equals to the CM-frame M\o ller velocity $v^{\rm CM}_\M$ defined in Eq.~\eqref{eq:flux}.}. Near the higgs pole the rate is dominated by quark and lepton final states, but contains also a non-negligible contribution from virtual gauge-boson final states. An accurate cross section can be obtained by using the factorizing into the $SSh$ fusion part times the virtual $h$ decay using the full width of the higgs~\cite{Cline:2012hg}:
\begin{equation}	
 v_\M^{\rm CM}\sigma_{\rm I} = 
         \frac{2\lambda_{\rm hs}^2 v_0^2 \,\Gamma_h(\sqrt{s})} 
	      {\sqrt{s}[(s-m_h^2)^2 + m_h^2\Gamma_h^2]}
\label{better}
\end{equation}
where $v_0=246$ GeV and the higgs decay width $\Gamma_h(\sqrt{s})$ is taken from ref.\ \cite{Dittmaier:2011ti}. 

\begin{figure}[t!]
\includegraphics[width=0.495\textwidth]{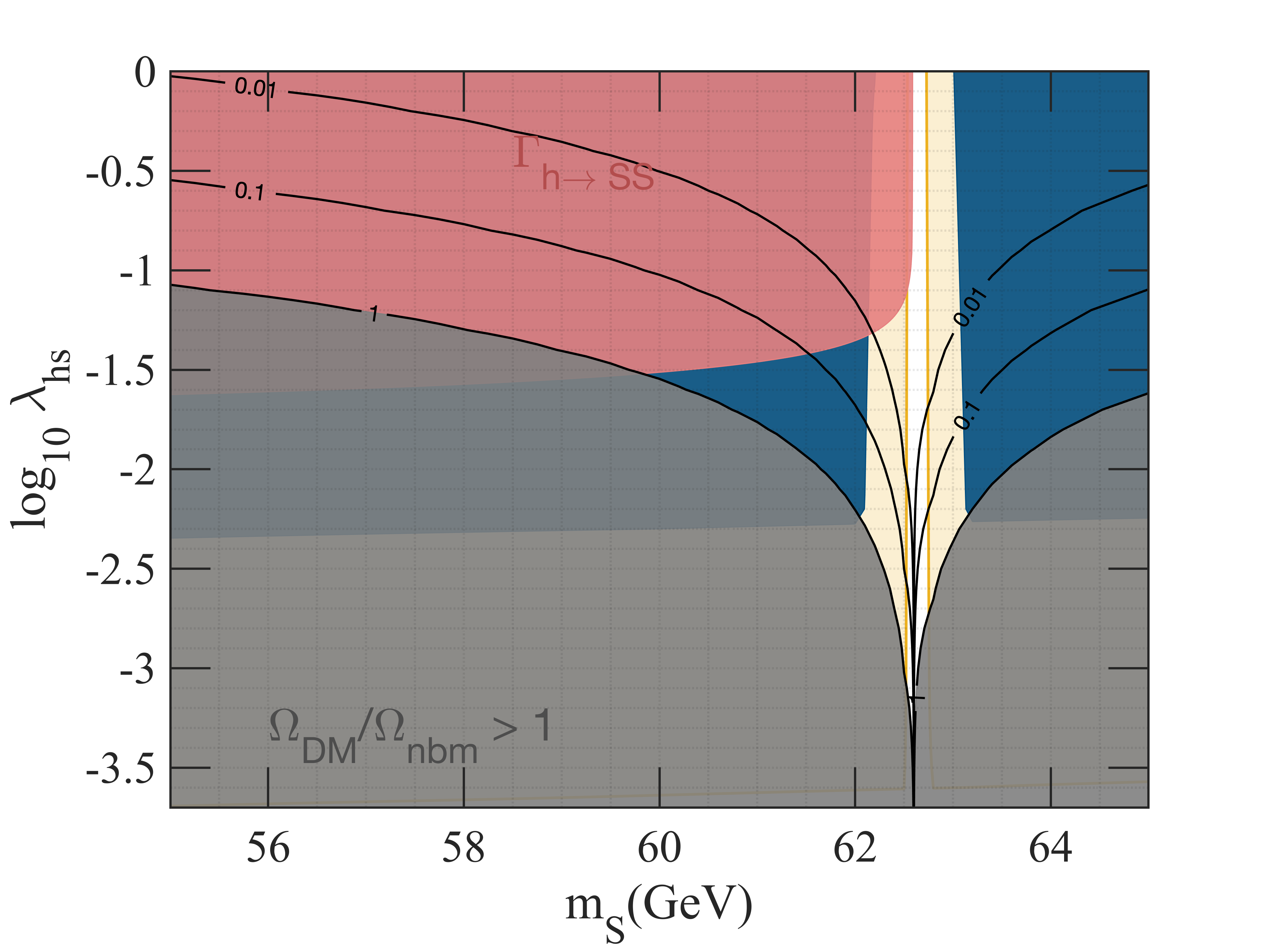}
\includegraphics[width=0.495\textwidth]{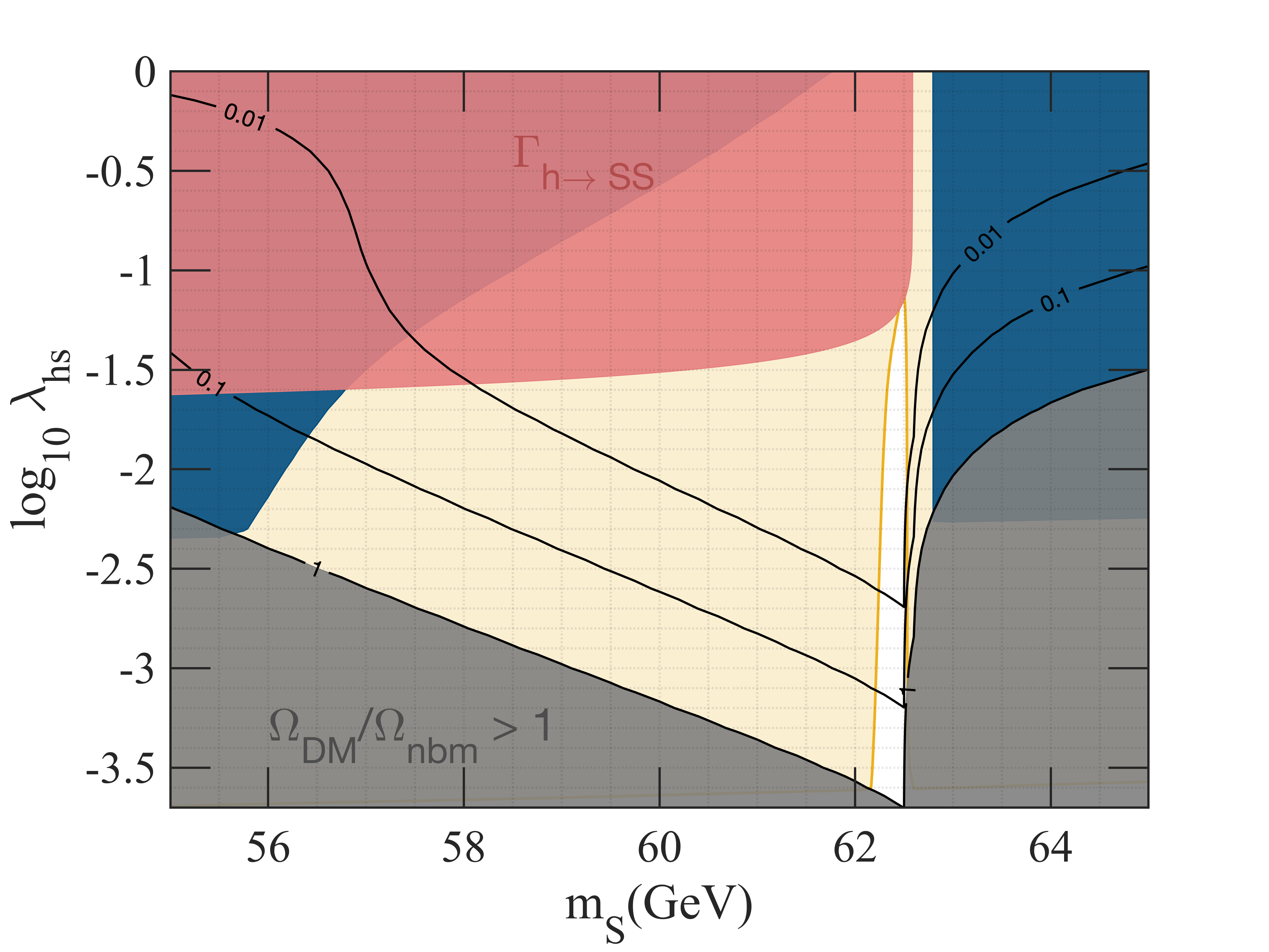}
\caption{Contours of fixed relic density as a fraction of the full dark matter density. \textit{Left}: Threshold cross section approximation. \textit{Right}: Thermal averaged cross section calculation. Gray areas are ruled out due to too large relic density. Dark blue areas are excluded by Xenon1t direct search limit~\cite{Aprile:2018dbl} and light yellow area shows the exclusion sensitivity of a hypothetical experiment reaching the neutrino floor. The red area is excluded by the higgs boson invisible width~\cite{Planck:2015xua}.}
\label{fig:MomAveContours}
\end{figure}

First, we find, in agreement with~\cite{Cline:2013gha}, that approximation (\ref{eq:Ytoday}-\ref{eq:xfout}) is consistent with the numerical integration of \eqref{eq:LWwithY} at one per cent level over the whole range considered. The relic abundance contours of a computation with full thermal cross section~(\ref{eq:Mollerintegral}) are shown in the right panel of figure~\ref{fig:MomAveContours}. In  this approximation the elastic collision integral vanishes and hence the results do not depend on $\lambda_\S$. Black curves show the contours of a constant relative dark matter density:
\begin{equation}
f_{\rm rel} \equiv \frac{\Omega_\S h^2}{0.1193} \,,
\end{equation}
where we used the latest CMB-determination for the the DM abundance $\Omega_\DM h^2 = 0.1193\pm 0.0014$~\cite{Planck:2015xua}. In the left panel we show the result of a computation which employs the threshold approximation~(\ref{eq:thresholdapp}) for $\langle v_\M\sigma_{\rm I} \rangle$. The difference is quite striking: the thermally averaged formula gives a much wider allowed region below the pole. We have shown both the current Xenon1t~\cite{Aprile:2018dbl} exclusion contour (dark blue) as well as the exclusion sensitivity of an hypothetical experiment reaching the sensitivity of the neutrino floor (yellow). 

The difference arises because $v_\M \sigma$ is a sharply peaked function of $s$ near the pole  and the threshold formula~(\ref{eq:thresholdapp}) does not account for kinetic energy of particles. Indeed, in finite temperature the kinetic energy of particles can make up for the missing mass and push collision energy to the pole\footnote{The effect of thermal averaging of the annihilation cross section near resonances was discussed quantitatively long time ago by Griest and Srednicki~\cite{Griest:1990kh}.}. Using Maxwell-Boltzmann statistics one finds that $\langle s\rangle \approx 4m^2 + 6mT$, and so one expects that thermally averaged cross section gets an asymmetric effective width below the pole of the order $\Gamma_{\rm eff} \sim 3m_h/4x_f\approx 4.3$ GeV, where we used $x_{\rm f} = m/T_{\rm f}\approx 22$. This simple argument indeed qualitatively explains the difference of the results shown in figure~\ref{fig:MomAveContours}.

%

However, it is not obvious that even the calculation using thermally averaged cross section can be trusted near the pole. The problem is that when the pole is very narrow, only particles with a finite range of momenta are sensitive to it. When these momenta are depleted, annihilations are less efficient until elastic interactions re-equilibrate the phase space.  Thus, while threshold approximation certainly overestimates the relic abundance, using full momentum averaged integral might well underestimate it. To see whether this really is so, one has to solve DM abundance using full momentum dependent Boltzmann equations.

%
\section{Momentum dependent problem}
\label{sec:MBapproximation}
%

\begin{figure}
\centering
\includegraphics[width=0.55\textwidth]{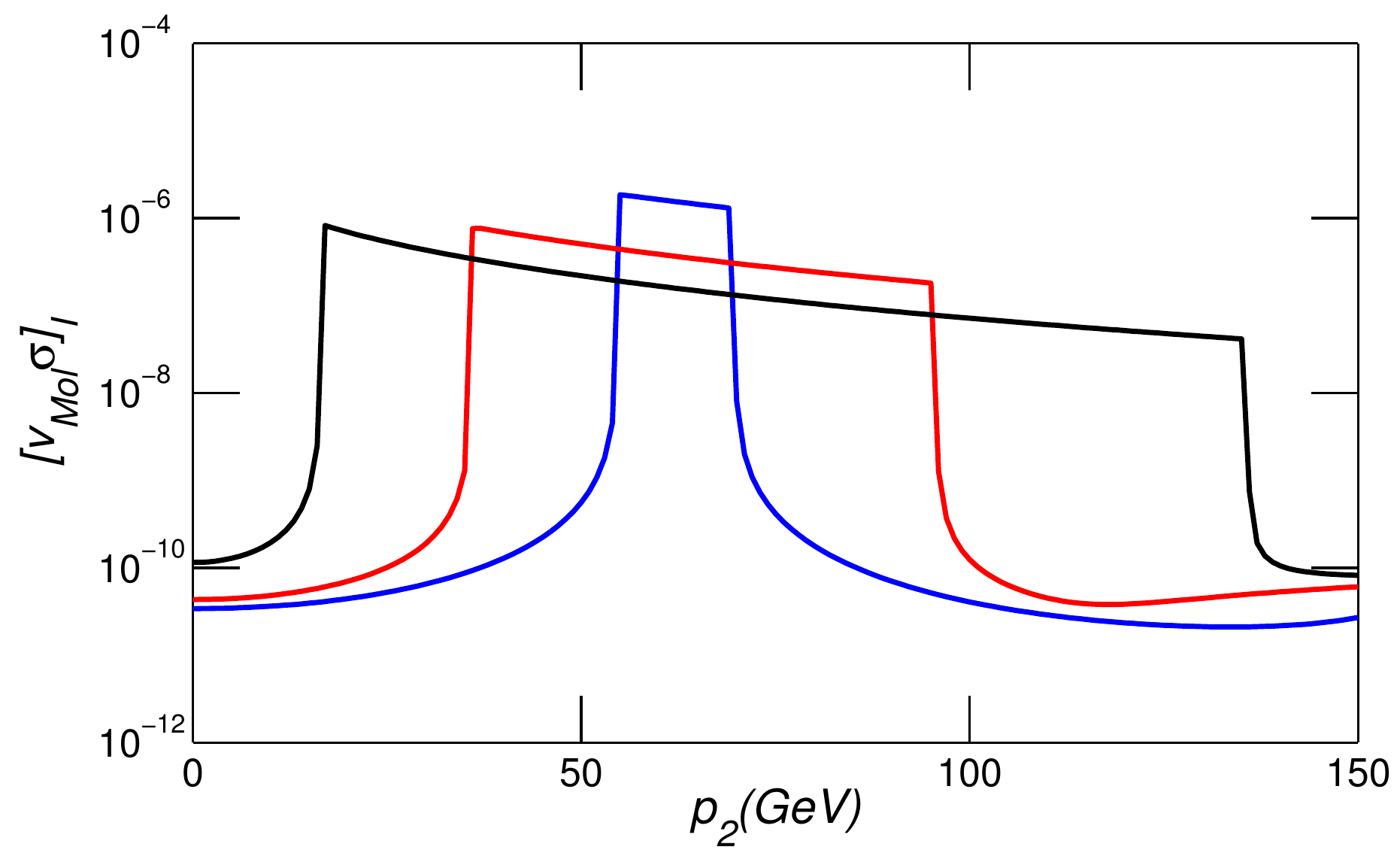}
\caption{$[v_\M\sigma]_{\rm I}$ of the singlet scalar model as a function of $p_2$ with $m_\S = 56$ GeV and $\lambda_{hs}= 0.01$ for $p=5$GeV, 20GeV and 40GeV (blue, red and black curves, respectively).}
\label{fig:vMoller1}
\end{figure}

We still assume that all SM particles involved in collisions are maintained in equilibrium at all times. We will also continue using Maxwell-Boltzmann statistics for equilibrium distribution functions. This is in fact a very good approximation when DM particles are non-relativistic and it brings great simplifications to collision integrals. Let us start by the inelastic collision integral. Given our assumptions, we can now write it as
\begin{equation}
\hat C^{\rm MB}_{\rm I}(p_1,t) 
= \frac{1}{2E_1} \int  \frac{{\rm d}^3 p_2}{(2\pi )^3 2E_2} 
  \left( e^{-\beta (E_{1}+E_{2})} - f(p_1,t) f(p_2,t) \; \right) 
  F_{\rm I}\sigma_{\rm I}(s)   \,,
\label{eq:Cinealstic2}
\end{equation}
where $\sigma_{\rm I}(s)=\sum_n \sigma_{\rm I,n}(s)$ is again the sum of inelastic cross sections to all available channels, and the flux-factor
\begin{equation}
F_{\rm I} = 4E_1 E_{2}v_\M = 2s \sqrt{1-\frac{4m_\S^2}{s}} \equiv 
sv_\M^{\rm CM} \,.
\label{eq:flux}
\end{equation}
One can always reduce the integral in (\ref{eq:Cinealstic2}) to one over the absolute value of the three momentum $p_2$ and $s$, which allows us to write:
\begin{equation}
\hat C^{\rm MB}_{\rm I}(p_1,t)  = 
f_{\rm eq}(p_1,t) \Gamma_{\rm I}[f_{\rm eq};p_1,t] - f(p_1,t) \Gamma_{\rm I}[f;p_1,t] \,.
\end{equation}
Here we wrote, for the sake of symmetry, $e^{-\beta E_{i}}=f_{\rm eq}(p_i,t)$ and defined
the decay functional:
\begin{equation}
\Gamma_{\rm I}[f;p_1,t] \equiv \frac{1}{2\pi^2} \int_0^\infty {\rm d}p_2p_2^2 \, f(p_2,t)
[v_\M \sigma_{\rm I}](p_1,p_2) \,,
\label{eq:gammaI}
\end{equation} 
where
\begin{equation}
[v_\M\sigma_{\rm I}](p_1,p_2) \equiv \frac{1}{s_+ - s_-}\int_{s_-}^{s_+}{\rm d}s \, v_\M\sigma_{\rm I}(s)   \,,
\label{eq:vmollersigmaI}
\end{equation}
with $s_\pm =  2m_\S^2 + 2E_1E_{2} \pm 2 p_1p_2$ . In practice, we can perform the s-integral in CM-frame:
\begin{equation}
    [v_\M\sigma_{\rm I}](p_1,p_2) = \frac{1}{16p_1p_2E_1 E_{2}} \int_{s_-}^{s_+} {\rm d}s \,s\, v_\M^{\rm CM} \, \sigma_{\rm I} (s)  \,.
\label{eq:vmollersigmaICM}
\end{equation}
The relevant $v_\M^{\rm CM} \, \sigma_{\rm I} (s)$ combinations for the SSM are given in ref.~\cite{Cline:2013gha}. The functional form of the quantity $[v_\M\sigma_{\rm I}](p_1,p_2)$ is relevant for the validity of the assumption of kinetic equilibrium: if this function is strongly peaked and elastic scatterings are weak, then the kinetic equilibrium assumption is not likely to hold. From Fig.~\ref{fig:vMoller1} we see that the situation is not disastrous: the $s$-averaging in Eq.~(\ref{eq:vmollersigmaI}) smooths the effect of the sharp peak in $\sigma_{\rm I}(s)$ considerably. There is, however, a huge enhancement for the momentum configurations that are sensitive to the pole (the flat top part in each graph), and so considerable momentum biases and changes in the final abundances may be expected to arise.
%

%
\subsection{Elastic collision term in relaxation approximation}
\label{sec:relax}
%

In the MB-approximation the elastic collision term between the scalar and SM particles immediately reduces to
\begin{eqnarray}
\hat C_{\rm E}(p_1,t) &\approx& \frac{1}{2E_1} \sum_m \int \left[ \prod_{i=2}^4
\frac{{\rm d}^3 p_i}{(2\pi )^32E_{i}}\right] (2\pi )^4\delta^4(p_1+p_3-p_2-p_4) \times
\nonumber \\
&&\times\mid {\cal M}_{13\rightarrow 24}^{(m)}\mid^2
                \Big(  f(p_2,t) f_{m}(p_4,t) - f(p_1,t)f_{m}(p_3,t) \Big)\,.
\label{eq:CelasticA}
\end{eqnarray}
Without further approximations, the best one can do is to reduce $C_{\rm E}(p_1,t)$ to a 5-dimensional integral, whose numerical evaluation would be very time-consuming. However, the problematic term (the first one in~(\ref{eq:CelasticA})) is a weighted integral over the target distribution function $f(p_2,t)$, whose precise shape is not crucial for the relaxation towards equilibrium.

It is therefore reasonable to make the following {\em generalized relaxation approximation}.  First, we continue to assume that all SM particles $m$ are in thermal equilibrium: $f_m \rightarrow f_{m,\rm eq}$. As a result, setting $f \rightarrow g f_{\rm eq}$ with an arbitrary function $g(t)$ makes elastic integral vanish. It then makes sense to factor the DM distribution as
\begin{equation}
f(p,t) = g(t) f_{\rm eq}(p,t) + \delta f(p,t) \,.
\label{eq:pseudoeddist}
\end{equation}
The term in brackets in (\ref{eq:CelasticA}) containing the distribution functions now becomes
\begin{equation}
  \delta f(p_2,t) e^{-\beta E_{4}}  - \delta f(p_1,t) e^{-\beta E_{3}} \,.
  \label{eq:Celastic2}
\end{equation}
As alluded above, the collision integral corresponding to the first term is a multi-dimensional convolution over the perturbation, which is typically a smooth function in $p$ even when $\delta f(p_2,t)$ itself is not a smooth function. The key element of our scheme is to use the freedom in choosing the function $g_m(t)$: we can in particular adjust it such that integrated elastic collision term corresponding to the division~\eqref{eq:Celastic2} vanishes {\em separately} for the forward and backward scattering terms. With this definition the back-reaction term should become a smooth, low amplitude variation around the actual elastic collision integral, whose integrated effect should be small. This term we then drop from our equation. We provide more details and an estimation of the accuracy of this approach by comparison to exact elastic collision integrals in the appendix \ref{sec:appC}. This corresponds to setting, separately for each elastic collision channel $m$:
\begin{eqnarray}
\hat C_{{\rm E},m}(p_1,t) &\rightarrow& 
- \delta f(p_1,t) \, \Gamma^{m}_{\rm E}(p_1,t)  \,  
\nonumber \\ 
&=& \left( \, g_{m}(t)f_{\rm eq}(p_1,t) - f(p_1,t)\, \right) \, \Gamma^{m}_{\rm E}(p_1,t)  \,,
\label{eq:Celast}
\end{eqnarray}
where $g_m(t)$ is defined to preserve the conservation of particle number in elastic collisions:
\begin{equation}
	\int\frac{{\rm d}^3p_1}{(2\pi)^3} \hat C_{{\rm E},m}(p_1,t) \equiv 0  \quad \Rightarrow \quad g_m(t) \equiv \frac{\int {\rm d}p_1 \, p_1^2\,f(p_1,t)\,\Gamma^m_{\rm E}(p_1,t)}{\int {\rm d}p_1 \, p_1^2\,f_{\rm eq}(p_1,t)\,\Gamma^{m}_{\rm E}(p_1,t)} \,.
\label{eq:gammam}
\end{equation}
The first term in the second line of the equation~\eqref{eq:Celast} replaces the the back-reaction term in the original elastic collision integral~\eqref{eq:CelasticA}. It ensures that $\hat C_{\rm E}(p_1,t)$ does not change the particle number and drives the distribution towards the pseudo-equilibrium form~\eqref{eq:pseudoeqdist}. Note that both equations~\eqref{eq:Celast} and~\eqref{eq:gammam} are essential: without the latter the former would make no sense.

After some manipulations each elastic rate function $\Gamma^{m}_{\rm E}(p_1,t)$ can be written in a similar manner as Eq.~(\ref{eq:gammaI}):
\begin{equation}
\Gamma^{m}_{\rm E}(p_1,t) \equiv  \Gamma^m_{\rm E}[f_{\rm eq}^{m};p_1,t] = 
  \frac{1}{2\pi^2}\int_{0}^\infty {\rm d}p_3p_3^2 \;     
    f^m_{\rm eq}(p_3,t) \,[v_\M\sigma ]^{\S m}_{\rm E}(p_1,p_3) \,,
\label{eq:gammaE}
\end{equation}
where we defined, similarly to Eq.~(\ref{eq:vmollersigmaI}):
\begin{align}
\label{eq:vmolsigmaEl}
[v_\M\sigma ]^{\S m}_{\rm E}(p_1,p_3) 
&= \frac{1}{8p_1p_3E_1E_3} \int_{s^m_-}^{s^m_+} {\rm d}s \, \lambda^{1/2}(s,m_{m}^2,m_\S^2) \, \sigma_{\rm E}^{\S m}(s)  \,.
\end{align}
Here $s^m_\pm =  m_m^2 + m_\S^2 + 2E_1E_{3} \pm 2 p_1p_3$ and $\sigma_{\rm E}^{\S m} (s)$ is the usual 2-body elastic cross section in channel $m$ and the kinetic function $\lambda(x,y,z)\equiv (x-y-z)^2 - 4yz$. 
Note that $m$, $\S$ and E are mere labels in equation~\eqref{eq:vmolsigmaEl}. This expression  is actually valid for any initial states $ab$, and both for the elastic and the inelastic interactions. In particular equation~\eqref{eq:vmollersigmaICM} is just a special case of~\eqref{eq:vmolsigmaEl}, where $ab = \S\S$ in the annihilation channel.

When applied to the case of self-scatterings of the scalar particles the above reasoning results to
\begin{eqnarray}
\hat C_{{\rm E},\S}(p_1,t) &\approx& 
g_\S^2(t)f_{\rm eq}(p_1,t)\,\Gamma^\S_{\rm E}[f_{\rm eq};p_1,t] - f(p_1,t)\,\Gamma^\S_{\rm E}[f;p_1,t]  \,,
\end{eqnarray}
where the decay function is defined in Eq.~\eqref{eq:gammaE} with the cross section $[v_\M\sigma ]^{\S\S}_{\rm E}$ and $g_\S$ is obtained from the conservation of particle number:
\begin{equation}
	\Rightarrow \quad g_\S^2(t) = \frac{\int {\rm d}p_1 \, p_1^2\,f(p_1,t)\,\Gamma^\S_{\rm E}[f;p_1,t]}{\int {\rm d}p_1 \, p_1^2\,f_{\rm eq}(p_1,t)\,\Gamma^\S_{\rm E}[f_{\rm eq};p_1,t]} \,.
\label{eq:conservation}
\end{equation}

%
\subsection{Boltzmann equation in co-moving momentum} 
%
The momentum derivative term $-Hp_1\partial_{p_1} f$ in the Liouville operator in equation~(\ref{eq:BoltzmannEq}) can be removed by taking the co-moving momentum $k_1 = p_1a$ as a new variable:
\begin{equation}
(\partial_t - Hp_1\partial_{p_1} ) f(p_1,t) = \partial_t \tilde f(k_1,t) 
\,,\phantom{\frac{2}{2}}
\end{equation}
where $\tilde f(k_1,t) \equiv f(p_1,t)$. The point is that $\tilde f(k_1,t)$ depends on $t$ only along the  characteristic lines of constant $k_1$. The time variable can then be traded for temperature just as we did in the momentum integrated case, assuming the adiabatic expansion $\dot s/s=-3H$. The relation between co-moving and physical momenta $k_1 = p_1a$ can then be read from ($a_0\equiv 1$):
\begin{equation}
 a = \left(\frac{s_0}{s}\right)^{1/3} = \left(\frac{h_{\rm eff}(T_0)}{h_{\rm eff}(T)}\right)^{1/3} \frac{T_0}{T}\,.
\label{eq:scalefactor}
\end{equation}
Combining the results, we can now write the full Boltzmann equations in the Maxwell-Boltzmann and relaxation approximations in terms of a dimensionless variable $x \equiv m_\S/T$ in the following simple form
\begin{eqnarray}
\partial_x \hat f(k_1,x) &=&  \hat f_{\rm eq}(k_1,x) X_{\rm I}[\hat f_{\rm eq};k_1,x] 
                        - \hat f(k,x) X_{\rm I}[\hat f;k,x] 
   \nonumber \\
    &+&g_\S^2(x)\hat f_{\rm eq}(k_1,x)X^\S_{{\rm E}}[\hat f_{\rm eq};k_1,x] - \hat f(k_1,x)X^\S_{{\rm E}}[\hat f;k_1,x]
   \nonumber  \\
     &+& \sum_{m}\left(\, g_m(x)\hat f_{\rm eq}(k_1,x) - \hat f(k_1,x) \,\right) X^{m}_{\rm E}(k_1,x)  \,,
\label{eq:BEinx}
\end{eqnarray}
where $\hat f(k_1,x) = \tilde f(k_1,t) = f(p_1,t)$, and 
\begin{equation}
X_{\rm i} \equiv 
\frac{1}{x}\left[1 + \frac{T}{3h_{\rm eff}}\frac{{\rm d}h_{\rm eff}}{{\rm d} T} \right] \frac{\Gamma_{\rm i}}{H}  \,,
\end{equation}
where $\Gamma_{\rm i}(p_1,t)$ are given by Eqs.~(\ref{eq:gammaI}) and (\ref{eq:gammaE}). As expected, for any given momentum variable $p_1$, the degree of equilibrium is defined by the ratio of the momentum-dependent interaction rate $\Gamma$ and the Hubble expansion rate $H$.

%
\subsection{Discretisation}
%

For numerical solution we need to discretise the momentum variables. This if formally quite simple. In discretised system integrals become simple matrix products. Let us now define a new dimensionless dependent variable as follows:
\begin{equation}
y_i(x)  \equiv  \frac{1}{2\pi^2 s_0} \Delta k_i k_i^2 \hat f(k_i,x) 
           = \frac{1}{2\pi^2 s} \Delta p_i p_i^2 f(p_i,x)\,.
\end{equation}
This is the actual differential number density in a given (co-moving) momentum bin divided by the (present) entropy density. The sum of the binned variables provide at any time an approximation for the integrated quantity $Y$:
\begin{equation}
\sum_i y_i  =  \frac{a^3n}{s_0} =  \frac{n}{s} = Y\,.
\end{equation}
In terms of $y_i$ the discretised Boltzmann equations become:
\begin{align}
\partial_x y_i =& - y_i\, \left(Z_{\rm I} \dy\right)_i 
+ y_{{\rm eq},i} \left(Z_{\rm I} \dy_{\rm eq} \right)_i 
\nonumber \\
& - y_i \left( Z_{\rm E}^{\S\S} \dy\right)_i + g_\S^2y_{{\rm eq},i} 
         \left(Z_{\rm E}^{\S\S} \dy_{\rm eq}\right)_i 
- \sum_{m} (y_i - g_my_{{\rm eq,i}}) Z^{m}_{{\rm E},i} 
\,,
\label{eq:finaleq1}
\end{align}
where $Z_{\rm I}$-term contains a sum over all available final states of equilibrium particles. The matrix products $(Z \dy)_i \equiv \sum_j Z_{ij}y_j$ replace one momentum integral each and $g_\f$ and $g_\S$ factors can be written simply as
\begin{equation}
    g_m = \frac{ \dy^{\rm T}  {\bf Z}^{m}_{\rm E}}{ \dy_{\rm eq}^{\rm T}  {\bf Z}^{m}_{\rm E}} \,,
\qquad {\rm and}\qquad
    g_\S^2 = \frac{ \dy^{\rm T}   Z_{{\rm E}}^{\S\S} \dy}
    { \dy_{\rm eq}^{\rm T}  Z_{{\rm E}}^{\S\S}  \dy_{\rm eq}}\,.
\end{equation}
Finally, the explicit forms of the discretised Z-functions are
\begin{equation}
Z^{ab}_{{\rm A},ij} \equiv 
\sqrt{\frac{\pi}{45}} g^{1/2}_*\; \frac{m_\S\MPl}{x^2} \; [v_\M \sigma]^{ab}_{\rm A} (p_i,p_j)\,, 
\label{eq:Zfinaali}
\end{equation}
where ${\rm A = I,E}$ and $[v_\M \sigma]^{ab}_{\rm A}(p_i,p_j)$ was defined in Eq.~\eqref{eq:vmolsigmaEl}. ${\bf Z}_{{\rm E}}^{m}$ can be computed directly using equation~\eqref{eq:trick2}, or during the integration of~\eqref{eq:finaleq1} from $Z_{{\rm E},i}^{m} = (Z_{{\rm E}}^{\S m}\dy_{\rm eq})_i$.
One should appreciate the similarity between equations (\ref{eq:finaleq1}-\ref{eq:Zfinaali}) with their integrated counterparts (\ref{eq:LWwithY}-\ref{eq:LWZ}). Indeed, (\ref{eq:finaleq1}) is but a set of coupled set of ZOPLW equations for the differential particle number elements with a elastic interactions providing a decay term towards the kinetic equilibrium. 

The two dimensional matrices $Z^{ab}$ need to be computed for each relevant interaction channel at each time step during the integration of~\eqref{eq:finaleq1}. Note however, that to compute them, we only need to know the {\em one-dimensional} integrals appearing in~\eqref{eq:vmolsigmaEl} for each cross section  (as a function of the upper limit, starting from $s=(m_a+m_b)^2$). These functions can be computed and fitted prior the integration, which speeds up the numerical integration tremendously.

%
\subsection{Generalisation to arbitrary number of species} 
%

It is straightforward to generalise our formalism to an arbitrary number of interacting species. If we denote these species by the set $\{A\}$, the equations take a very simple form:
\begin{equation}
\partial_x y^a_i = - y_i^a\, \big(Z^{ab} \dy^b \big)_i + G^{ab}_{cd}\;y_{{\rm eq},i}^a \big(Z^{ab} \dy_{\rm eq}^b \big)_i 
\,,
\label{eq:genSahaEq}
\end{equation}
where ${a,b,c,d\in A}$ are flavour indices for particles involved in the scattering process $ab \rightarrow cd$, and a sum over all allowed channels $b,c,d$ is assumed for each $a$. Here $x \equiv m/T$ where $m$ is some arbitrary reference mass. The "generalised Saha-factor" is given by\footnote{If we sum equation~(\ref{eq:genSahaEq}) over momenta and use the kinetic equilibrium approximation~(\ref{eq:pseudoeqdist}), which allows to write $\dy^e Z^{ab} \dy^f = \langle Z^{ab}\rangle_{\rm eq} Y^eY^f$, the equation~(\ref{eq:genSahaEq}) takes the form: $\partial_x Y^a = \langle Z^{ab}\rangle_{\rm eq} (Y^aY^b  - {\bar G}^{ab}_{cd}Y^cY^d)$, where ${\bar G}^{ab}_{cd}= (Y_{\rm eq}^aY_{\rm eq}^b)/(Y_{\rm eq}^cY_{\rm eq}^d)$ is the usual Saha factor in averaged momentum equations.}
\begin{equation}
    G^{ab}_{cd} \equiv  \frac{(\dy^c)^{\rm T}  Z^{ab}  \dy^d }{(\dy^c_{{\rm eq}})^{\rm T}  Z^{ab}  \dy^d_{\rm eq} }\,.
\label{eq:sahafactor}
\end{equation}
Note that here we have used explicit indices only to indicate a nontrivial dependence on distribution functions. Of course for example $Z^{ab}_{ij}$ depends on the species $c,d$ through the cross section.

Equation~\eqref{eq:finaleq1} can be obtained from~\eqref{eq:genSahaEq} with the following assignments: let $M$ be the subset of particles in $A$, which are in equilibrium with the SM heat bath and $m \in M$ (because $\dy^m=\dy_{\rm eq}^m$ we do not need an equation for $m$). Now the inelastic scattering term in Eq.~\eqref{eq:finaleq1} is recovered by setting $(a,b,c,d)\rightarrow(s,s,m,m)$, and noting that $m$ is in equilibrium, whereby:
\begin{equation}
    \Rightarrow \quad G^{\S\S}_{mm} = 1  \,.
\end{equation}
We get the elastic self-scattering term by setting  $(a,b,c,d)\rightarrow(s,s,s,s)$:
\begin{equation}
    \Rightarrow \quad G^{\S\S}_{\S\S} = \frac{(\dy^\S)^{\rm T}  Z^{\S\S}  \dy^\S }{(\dy^\S_{{\rm eq}})^{\rm T}  Z^{\S\S}  \dy^\S_{\rm eq} }
    = g_\S^2\,,
\end{equation}
and the elastic scattering with equilibrium particles by setting $(a,b,c,d)\rightarrow(s,m,s,m)$:
\begin{equation}
    \Rightarrow \quad  G^{\S m}_{\S m} \rightarrow \frac{(\dy^\S)^{\rm T}  Z^{\S m}  \dy^m_{\rm eq} }{(\dy^\S_{{\rm eq}})^{\rm T}  Z^{\S m}  \dy^m_{\rm eq} }
    = \frac{(\dy^\S)^T {\bf Z}^{m}_{\rm E}}{(\dy^\S_{{\rm eq}})^T  {\bf Z}^m_{\rm E} }
    = g_m \,,
\end{equation}
The generalised equation \eqref{eq:genSahaEq} is necessary when one has a more complicated Dark Sector consisting of at least two new particles, relatively closely spaced in mass. It could also be easily adapted to study novel out-of-equilibrium particle processes during nucleosynthesis, or for an accurate solution of particle distributions interacting with the expanding electroweak phase transition wall. In this paper we shall restrict ourselves to the simple example of a singlet scalar dark matter model.

%
\section{Numerical results} 
\label{sec:numerical_results}
%

We now present numerical comparisons of the dark matter abundance calculations. In the SSM the singlet can can be either a thermal WIMP, the case we have been studied so far, or it can be a feebly interacting massive particle (FIMP). We shall consider these two cases separately, starting from the thermal DM scenario near the resonance.

%
\subsection{Thermal DM} 
\label{sec:results_wimp}
%

\begin{figure}[t!]
\includegraphics[width=0.495\textwidth]{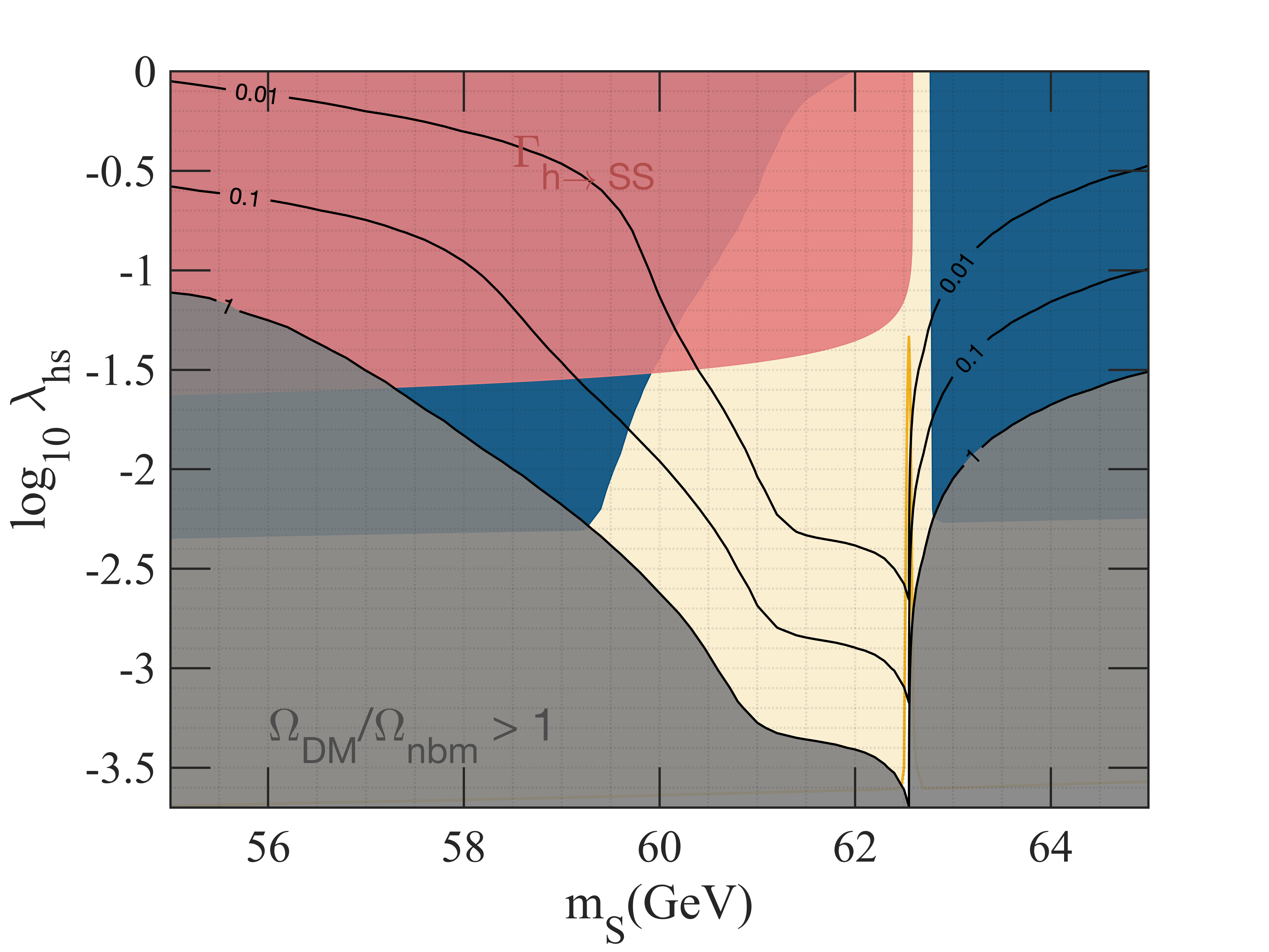}
\includegraphics[width=0.495\textwidth]{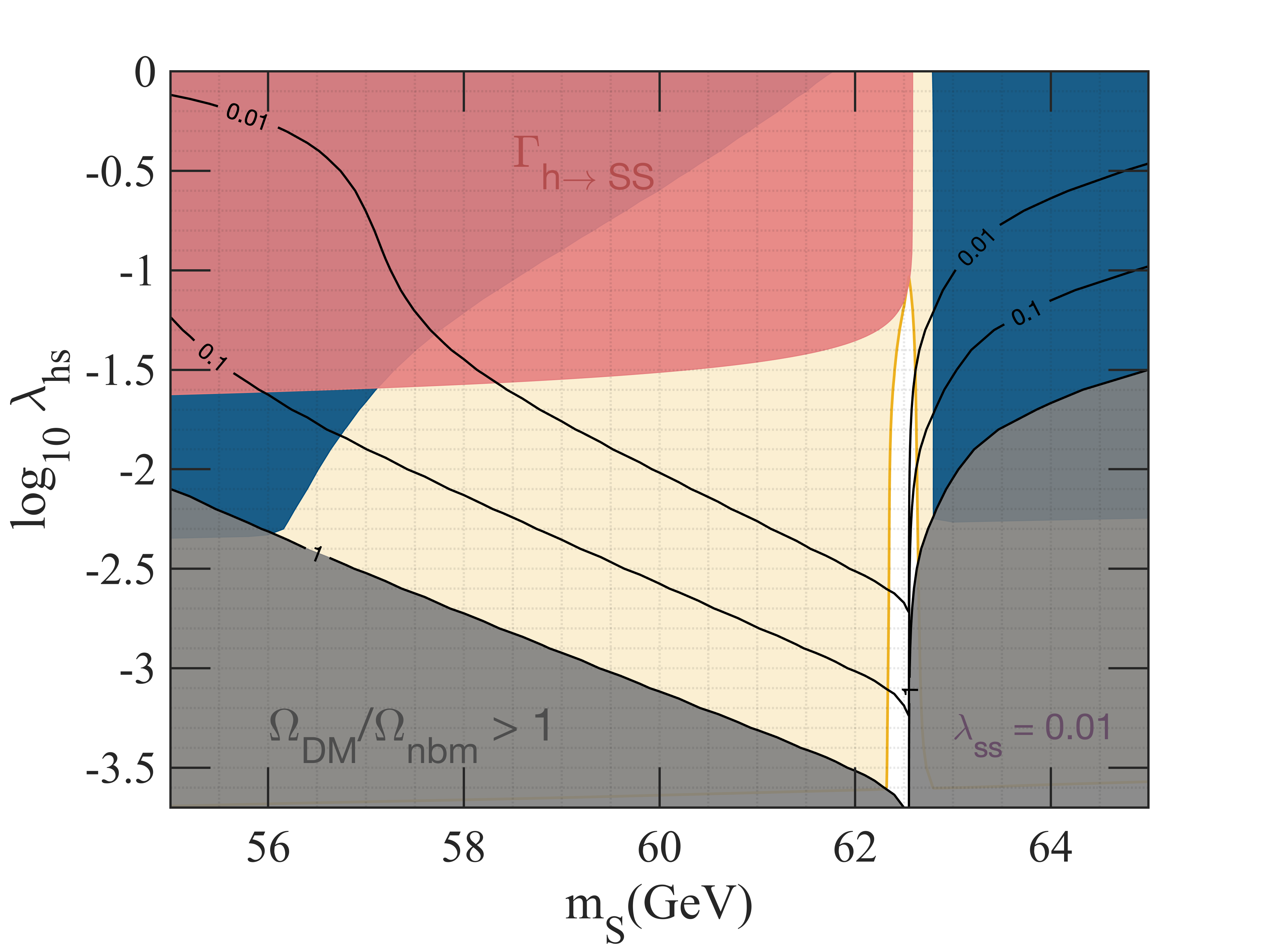}
\caption{Contours of fixed relic density as a fraction of the full dark matter density obtained from the momentum-dependent solution. \textit{Left}: Calculation with only the inelastic processes $ss\leftrightarrow \overline{\{SM\}}\{SM\}$. \textit{Right}: Calculation with the inelastic processes and the self-scattering process $ss\leftrightarrow ss$. The meaning of the various coloured contours and lines are as in figure~\ref{fig:MomAveContours}.}
\label{fig:BE1}
\end{figure}

In order to elaborate the effect of elastic scatterings on the abundance we present the momentum dependent calculation in various different approximations. First, we include only the inelastic scattering terms in the equations~\eqref{eq:finaleq1}. The relevant cross section needed to compute $Z_{I,ij}$ from~\eqref{eq:Zfinaali} using~\eqref{eq:vmollersigmaICM} is given in equation~\eqref{spot}. We show the result of the calculation in the left panel of figure~\ref{fig:BE1}. Not surprisingly, we find significantly higher abundances than we did earlier under the kinetic equilibrium assumption (right panel of figure~\ref{fig:MomAveContours}); the difference comes from the expected depletion of the states amenable for resonant scattering. Note that not only the equal abundance contours, but also the direct search exclusion limits change significantly in going from one approximation to another in Figs.~\ref{fig:MomAveContours} and \ref{fig:BE1}.

In the right panel of figure~\ref{fig:BE1} we show results with complete set of elastic interactions including the self-interactions $ss\leftrightarrow ss$ and the scatterings with the standard model particles (labelled by $m$) $sm\leftrightarrow sm$. The relevant scattering rates needed to compute $Z_{\rm E}^{\S\S}$ and $Z_{\rm E}^m$ from~\eqref{eq:Zfinaali} using~\eqref{eq:vmolsigmaEl} are given in the appendix~(\ref{eq:sigmaElf}-\ref{eq:sigmaEls}). We used the following fermion masses: $m_s = 95$ MeV, $m_c = 1.27$ GeV, $m_b = 4.18$ GeV and $m_\mu = 105.7$ MeV and $m_\tau = 1.777$ GeV, while other fermions were taken to be massless. For the $s$-self coupling we used  $\lambda_s = 0.01$. We can infer from figure~\ref{fig:BE1} that the current extent of the DM mass in the SSM is $m_\S \in [56,62.5]$ GeV. Moreover wee see that DM would be discoverable over this whole range in a direct DM search experiment whose sensitivity only slightly exceeds the neutrino floor.\\

Obviously our final results with full elastic interactions are almost identical with the kinetic equilibrium case shown in right panel of figure~\ref{fig:MomAveContours}. This result calls for some discussion. First, it shows that using ZOPLW equations with the thermally averaged cross section~\ref{eq:Mollerintegral}, gives the DM abundance accurately even below the sharp resonance in higgs portal models. This level of accuracy is easily sufficient for any exploratory research in the dark matter problem. However, one may ask if this is a generic feature, or just a particular property of the SSM? Indeed, what interactions were mostly responsible for achieving the kinetic equilibrium?

\begin{figure}[t!]
\begin{center}
\includegraphics[width=0.495\textwidth]{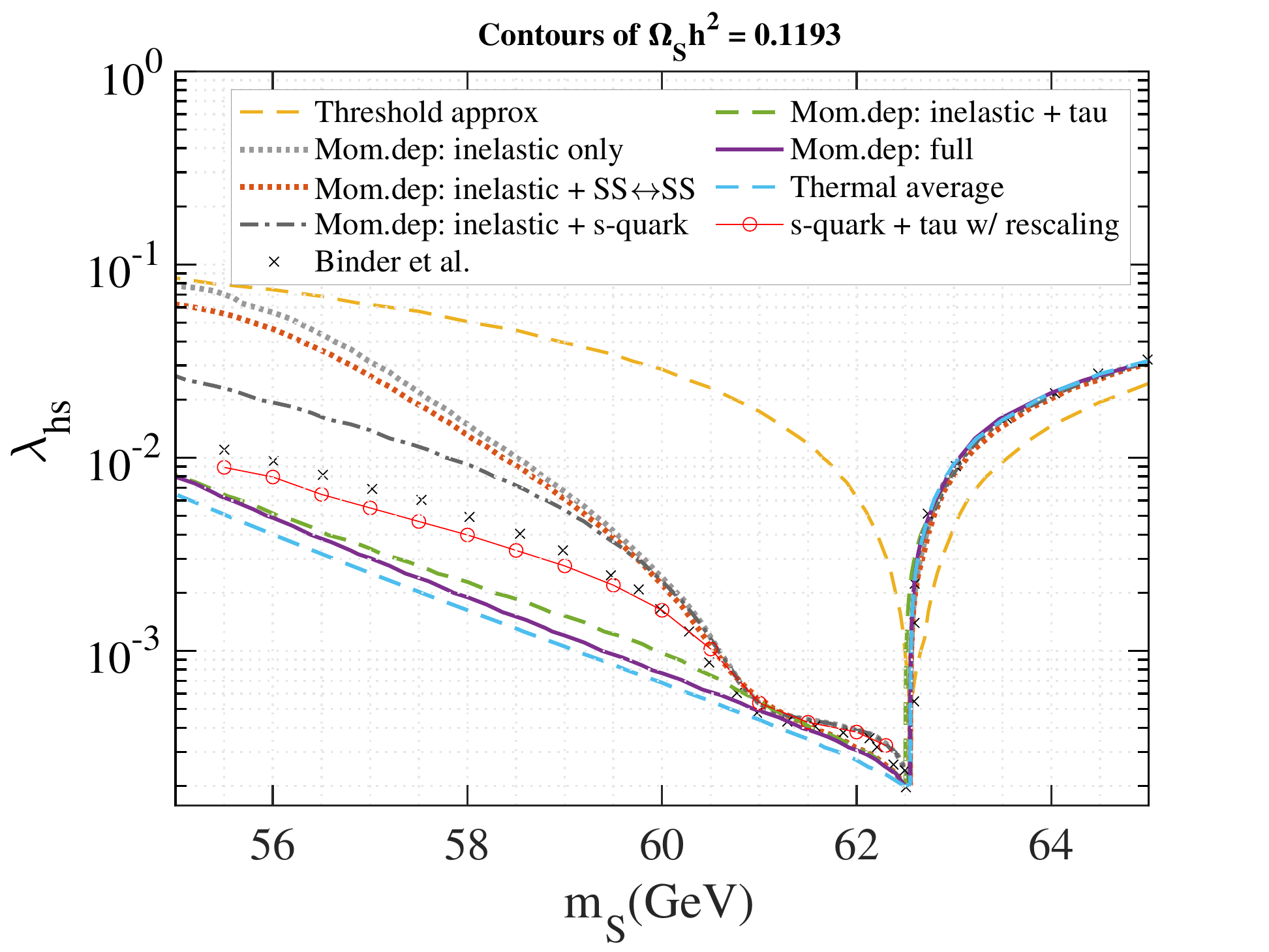} 
\includegraphics[width=0.495\textwidth]{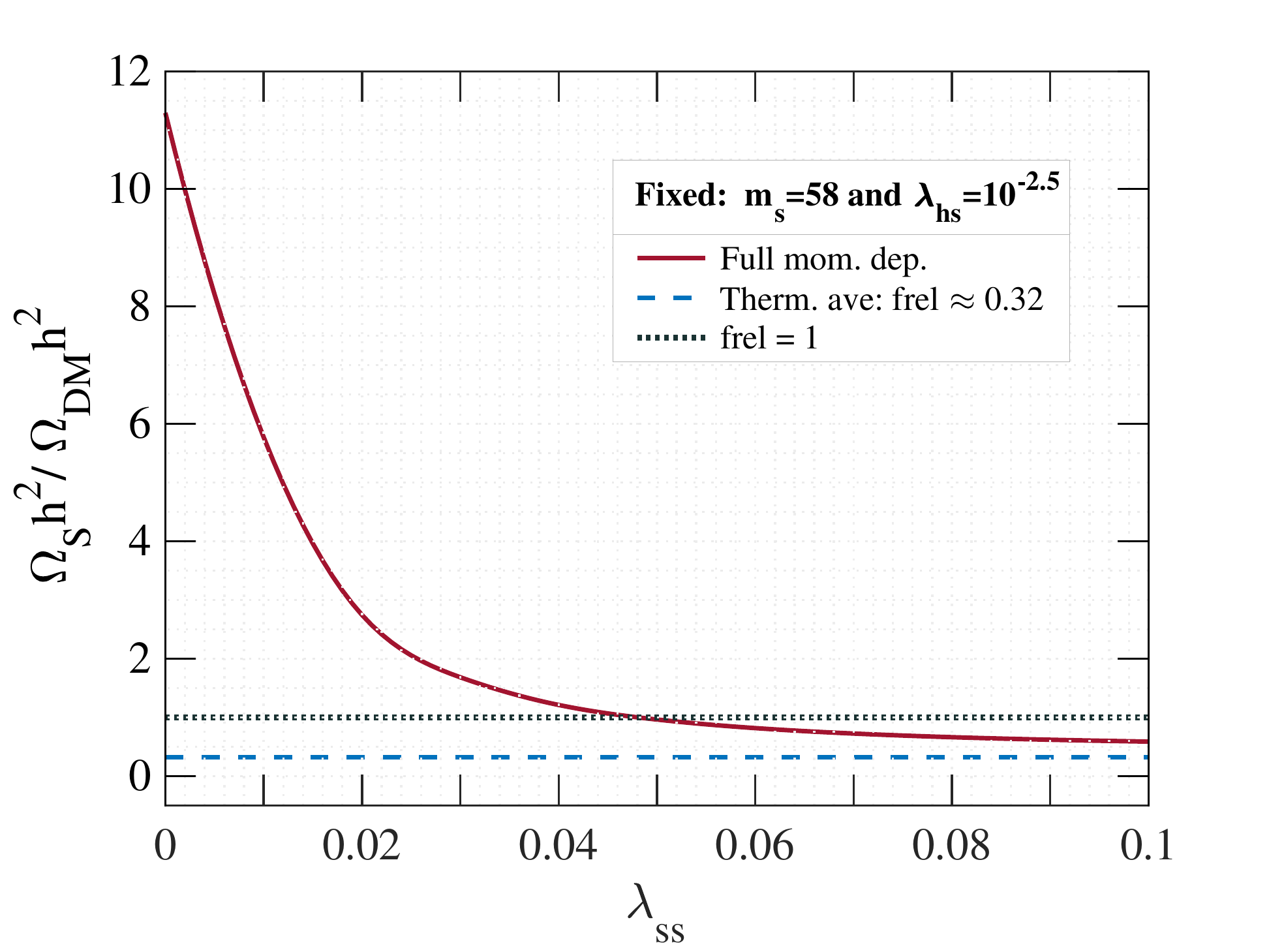}
\caption{\textit{Left}: Contours of $f_{\rm rel} = 1$ in various different assumptions for elastic interactions (see text). \textit{Right}: $f_{\rm rel}$ as a function of $\lambda_\S$ for $m_\S = 58$ GeV and $\lambda_{hs} = 10^{-2.7}$, with only the self-coupling induced elastic interactions.}
\label{fig:BE2}
\end{center}
\end{figure}

To study these questions we performed the analysis for several restricted sets of elastic interactions and the results are shown in the left panel of the figure~\ref{fig:BE2}. All lines displayed here show the $f_{\rm rel}=1$ contour in the approximation used. The light blue dashed line is the kinetic equilibrium result and the yellow dashed line corresponds to using ZOPLW equation in the threshold approximation. All other contours correspond to momentum dependent calculations: the purple solid line corresponds to full elastic interactions and the the gray dotted line shows the result with no elastic interactions. Almost overlapping with the latter, the red dotted line present the case with the elastic self-interactions only, again with $\lambda_s = 0.01$. Clearly, elastic interactions with the SM-states alone are sufficient for establishing the kinetic equilibrium.

The main SM contributions to the elastic scattering come from bottom and charm quarks and tau leptons and to lesser extent from strange quarks and muons. Contributions from all other fermions are negligible. Indeed, a typical freeze-out temperature, calculable from~\eqref{eq:xfout} is $T_f \approx m_h/(2x_f) \approx 3$ GeV.  This is well above the QCD phase transition and yet low enough such that only the $b$-quark population is slightly suppressed at the freeze-out. Somewhat surprisingly, including only tau leptons already almost saturates the equilibrium limit. We show the result of this calculation by the green dashed line in figure~\ref{fig:BE2}.

Including the charm and bottom quark contributions can change the result only slightly. If we fix the mass and coupling as $m_\S=58$ GeV and $\lambda_{\rm hs}=10^{-2.7}$, we find 
$f_{\rm rel} = 1.3$ with tau-channel only and $f_{\rm rel} = 0.85$ with full elastic interactions. Finally, kinetic equilibrium calculation gives $f_{\rm rel} = 0.7$. There thus remains a 20 per cent difference in results even with the full elastic scattering strength. Given a positive identification of the dark matter particle and high accuracy measurement of its properties, the momentum dependent calculation could still be necessary to establish consistency with the DM abundance.

One might wonder if the remaining difference could in principle be used to obtain information from the self coupling $\lambda_\S$? This appears not the case however; we find that varying $\lambda_\S$ in the range $[0,2\pi]$ changes $f_{\rm rel}$ by less than one per cent in the case with the full elastic SM-interactions. This is understandable because $\lambda_\S$ can induce equilibrium with the SM heat bath only indirectly, together with the inelastic rate. It is the inefficiency of the latter that produces the bottleneck for this equilibration mechanism.

For comparison we show in the right panel of figure~\ref{fig:BE2} the effect of $\lambda_\S$ excluding all elastic SM-scatterings. In this case $\lambda_\S$ has a strong effect. A coupling of order $\lambda_\S \gsim 0.07$ is sufficient to establish a reasonably complete kinetic equilibrium. This case may be representative of more complicated models, where DM is not necessarily directly coupled with SM.

%
\
%

\subsubsection{Comparison to earlier work}

SSM was recently analysed using both moment expansion and momentum dependent Boltzmann equations in~\cite{Binder:2017rgn}, with results that are qualitatively similar to ours. In particular ref.~\cite{Binder:2017rgn} found that elastic scatterings with quarks may enforce the kinetic equilibrium. However, they also concluded that the correct DM abundance in SSM can differ by an order of magnitude from the one found by traditional treatment, depending on the characteristics of the QCD phase transition. We do not find any such dependence here. Instead, all our conclusions are, as explained, robust against any assumptions about QCD. Most of the discrepancy appears to stem from an error in ref.~\cite{Binder:2017rgn} equation (42), which underestimates the matrix element squared for elastic scalar-fermion scatterings by a factor of 8.

Indeed, the scenario B of~\cite{Binder:2017rgn} should correspond to our case including only strange quark and lepton elastic scatterings, but their results (shown by crosses in figure~\ref{fig:BE2}) differ from ours by a factor up to 2 in coupling. Artificially reducing our elastic rates by a factor 8 in this case gives the thin red line with red circles. The remaining difference is qualitatively consistent with the different approximations to the elastic collision integrals, which in~\cite{Binder:2017rgn} were computed in (semi-) relativistic expansions, in a zero momentum transfer approximation. While our relaxation time scheme tends to slightly overestimate the elastic integrals, the method of ref.~\cite{Binder:2017rgn} tends to underestimate them. See appendix~\ref{sec:appC} for a detailed comparison of these approximations against exact collision integrals.

Our analysis also contains features not included in ref.~\cite{Binder:2017rgn}, such as the role of the self-scatterings as well as the computation of the direct detection constraints. Our formalism is also more transparent and valid for arbitrary number of interacting species. Ref.~\cite{Binder:2017rgn} also only considered the thermal WIMP case, whereas we also study the possibility of a feebly interacting dark matter in the SSM.

%
\subsection{The FIMP scenario} 
\label{sec:results_fimp}

The SSM model allows also for another type of dark matter, a feebly interacting massive particle (FIMP). We saw above that going to smaller couplings in the WIMP region eventually leads to the DM overproduction. However, when the coupling is small enough, the DM may never be thermalised, which avoids this outcome. In the FIMP mechanism (for a review see~\cite{Bernal:2017kxu}), the coupling $\lambda_{\rm hs}$ is adjusted such that DM is only partly brought into equilibrium, giving just the desired DM abundance.

\begin{figure}[t!]
\begin{center}
\includegraphics[width=0.49\textwidth]{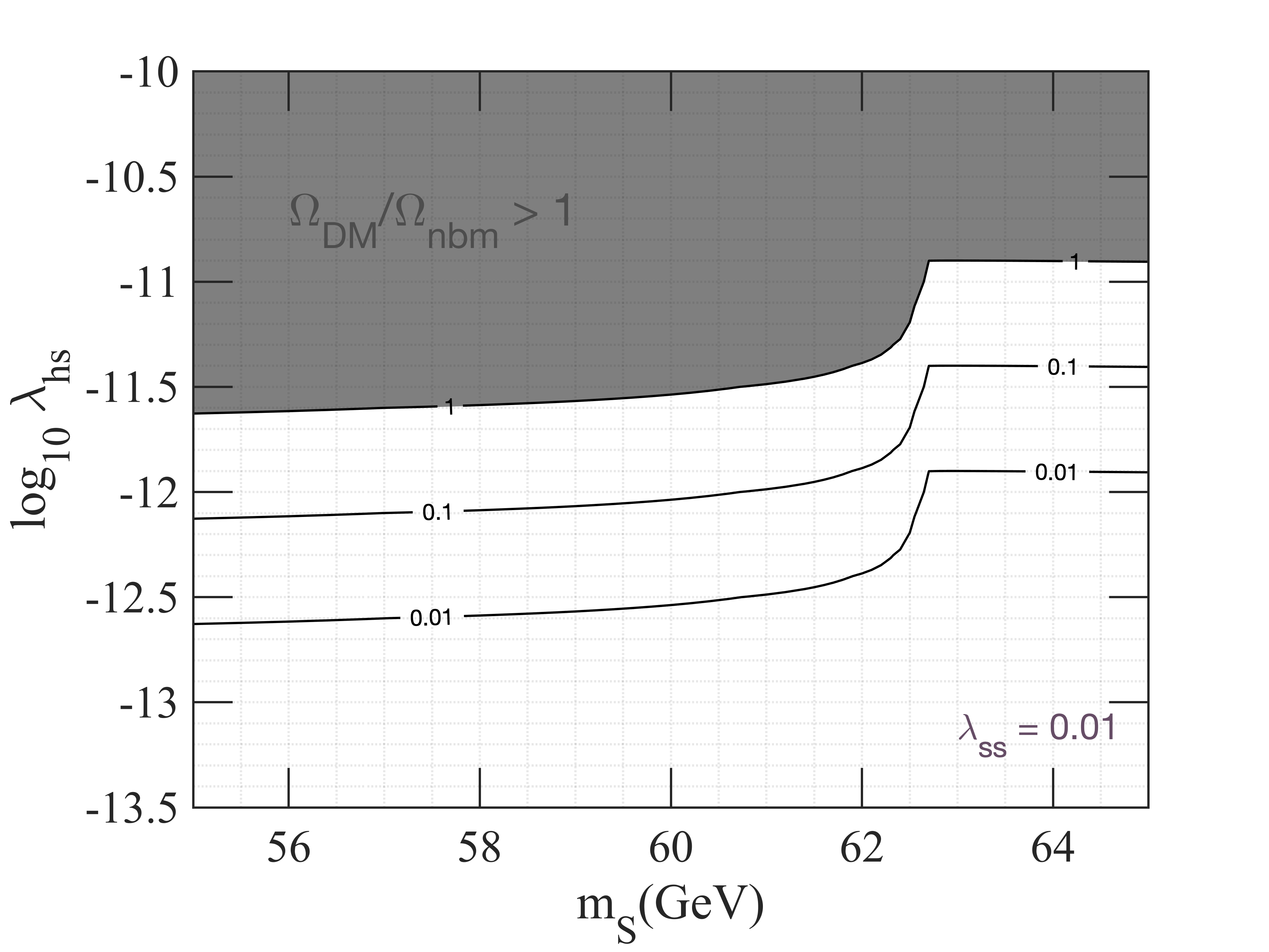} 
\includegraphics[width=0.49\textwidth]{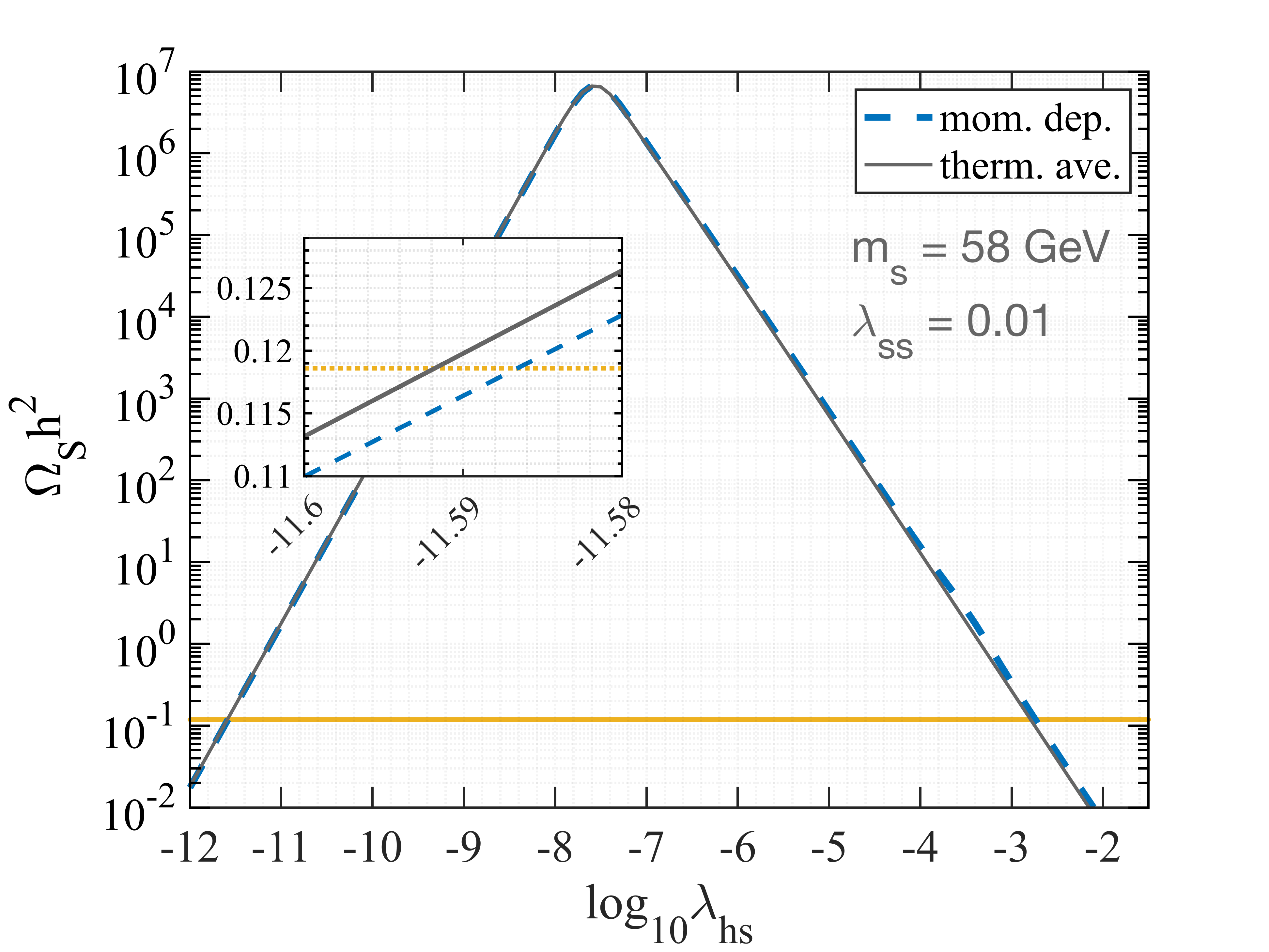}
\caption{Left panel: constant $f_{\rm rel}$ contours in the FIMP case calculated from the full momentum dependent code.
Right panel: shown is the dependence of the FIMP abundance as a function of $\lambda_{\rm hs}$ for the case $m_s = 58$ GeV and $\lambda_s = 0.01$. The gray line corresponds to the momentum averaged code calculation and the blue dashed line to the full momentum dependent calculation.}
\label{fig:fimp_figure_2}
\end{center}
\end{figure}

In the left panel of figure~\ref{fig:fimp_figure_2} we show the contours of constant $f_{\rm rel}$ in the FIMP region given by our full momentum dependent code\footnote{Note that in the FIMP case the MB approximation assumed by our method is not as robust as for WIMPs. In the resonant region the statistics corrections are expected to be only a few per cent, but above the resonance, where the FIMP production is dominated by the $W$ and $Z$ initial states, our neglect of the Bose-statistics factors can underestimate the abundance by a factor of two~\cite{Belanger:2018ccd}.}. The shape of these contours differ significantly from those in the WIMP region. In the WIMP case the abundance is determined in a narrow temperature range near freeze-out, whereby $f_{\rm rel}$ inherits the characteristic shape of the inverted annihilation rate. In the FIMP case the DM production occurs at much higher temperatures and all FIMPs with $2m_\S < m_h$ are produced resonantly at some point. On the other hand, FIMPs with $2m_\S > m_h$ are never sensitive to the pole. As a result, the effect of the pole does not show up as an inverted peak, but as a step-like structure at $2m_\S \approx m_h$. Our results agree qualitatively with ref.~\cite{Bernal:2018kcw}. 

%
%

Finally, in the right panel of figure~\ref{fig:fimp_figure_2} we show a comparison of the 
the FIMP abundance computed using the momentum averaged code (gray line) and the full momentum dependent code (dashed blue line) for representative parameters. For a given $\lambda_{\rm hs}$, the results agree to within 20 per cent. This agreement is expected, since FIMPs are produced gradually from the SM heat bath, subject to continuous elastic scatterings with the SM particles.

Of course SSM is but an example of a portal dark matter. More elaborate portal models and models with larger dark sectors and different types of the dark matter have been discussed in literature~\cite{Merle:2015oja,Bernal:2015xba,Konig:2016dzg,Heikinheimo:2017ofk,Heikinheimo:2018esa}, in many of which the DM would be expected not to be in thermal equilibrium. In some cases the shape of the non-thermal DM distribution may have an effect on observable quantities~\cite{Merle:2015oja,Heikinheimo:2016yds,Konig:2016dzg,Murgia:2017lwo,Murgia:2018now}. In such cases the momentum averaged methods are of course completely inadequate.
Our momentum dependent method would be easily implemented in all these studies to obtain most accurate results.

%

\section{Conclusions} 
\label{sec:conclusions}

We have presented a careful analysis of dark matter abundances using different approaches from analytic approximations to novel numerical momentum dependent methods. In particular we focused on the DM problem near sharp resonances, which appear for example in popular higgs portal models. We used the singlet scalar model (SSM) as a prototype and found that the momentum averaged approach based on the kinetic equilibrium approximation works very well even near sharp resonances. We updated the extent of the currently allowed region in the light singlet scalar dark matter to $m_\S \in [56,62.5]$ GeV. We also showed that the light DM in the SSM would be discoverable in a direct detection experiment whose sensitivity reach only slightly exceeds the neutrino floor.

In the SSM the residual error of using momentum averaged method is only 20-30 per cent. The result is robust and, unlike stated in ref.~\cite{Binder:2017rgn}, not sensitive on details of the QCD phase transition. We point out that even this deviation could be large enough to necessitate the use of momentum dependent methods for consistent results if DM particle was eventually observed and its mass and interaction strength were measured with very high accuracy.

In the SSM the kinetic equilibrium is mainly established by the elastic scatterings with the SM particles. The self-scatterings play no relevant role and the DM abundance cannot be used to constrain the SSM self-coupling $\lambda_\S$. However, there are other DM frameworks which may have suppressed elastic scatterings with the SM. In such cases self interactions would have a crucial role in restoring the kinetic equilibrium.

As a by-product of our analysis, we developed a very simple and generic numerical scheme for solving momentum dependent Boltzmann equations. The novel element of our scheme is the use of a generalised relaxation approximation to write the back-reaction collision integrals in terms of equilibrium quantities multiplied by simple Saha-like factors. All collision terms are reduced to universal one-dimensional integrals over the relevant CM-frame cross sections. The final equation~\eqref{eq:genSahaEq} is one of the main results of this paper. This formulation of the Boltzmann equations should be useful also in other out-of-equilibrium systems, such as the plasma interacting with the expanding electroweak phase transition walls.

Returning to the DM problem, we also studied the FIMP region in the SSM. Also here we found that the FIMP production takes place in a very near kinetic equilibrium and momentum averaged method is accurate to within 20 per cent. We finally point out that in more elaborate DM models with larger dark sectors the DM might not be in thermal equlibrium. Our momentum dependent method would be easily implemented in these studies as well.

%

%
\section*{Acknowledgements}
We thank Matti Heikinheimo and Kimmo Tuominen for discussions and comments. This work was supported by the Academy of Finland grants 310130 and 318319. We thank anonymous referee for pushing us to sharpen our comparison to ref.~\cite{Binder:2017rgn} and to provide the quantitative error analysis presented in the appendix~\ref{sec:appC}.
%

\FloatBarrier
\appendix

%
\section{Appendix:  CM-frame cross sections}
\label{app:exact_x-sections}
%

Elastic cross section for scalar-fermion collision is:

\begin{equation}
\sigma_{\rm E,f}(s) = \frac{N_{\rm f} \, \lambda_{\rm hs}^2\, m_{\rm f}^2}{4\pi\,\lambda(s,m_{\rm f}^2,m_\S^2)}
\left( \,\frac{(4m_{\rm f}^2 - m_\h^2)\,\lambda(m_\S^2,m_{\rm f}^2,s)}{m_\h^2\left(s\,m_\h^2 + \,\lambda(m_\S^2,m_{\rm f}^2,s)\right)}
+ \log\left(1 + \frac{\lambda(m_\S^2,m_{\rm f}^2,s)}{s\,m_\h^2} \right) 
\,\right)\,,
\label{eq:sigmaElf}
\end{equation}
where $\rm f$ denotes any SM-fermion. The cross section for the scalar self-scattering is:
\begin{align}
\sigma_{{\rm E},s}(s) = \frac{1}{32\pi\,s}
\Bigg[ \,
\left|a\right|^2&
+ \frac{2\lambda_{\rm hs}^4\,v^4}{m_\h^2(s+m_\h^2-4m_\S^2)} + \nonumber \\ 
+ &\frac{4\lambda_{\rm hs}^2\,v^2}{(s-4m_\S^2)}
\left(\,{\rm Re}(a) + \frac{\lambda_{\rm hs}^2\,v^2}{(4m_\S^2 - s -2m_\h^2)} \,\right)
\log\left| \frac{m_\h^2}{s+m_\h^2-4m_\S^2} \right| 
\,\Bigg]\,,
\label{eq:sigmaEls}
\end{align}
with 
\begin{equation}
    a \equiv 6\lambda_{\S} + \dfrac{\lambda_{\rm hs}^2\,v^2}{s-m_\h^2+i\sqrt{s}\Gamma_\h}\,,
\end{equation}
where $\Gamma_\h$ is the total higgs width, including the invisible contribution due to $h \rightarrow SS$ for $m_\S < m_\h/2$ region and $\lambda_\S$ is the 4-point self-coupling constant. For the higgs field vacuum expectation value we used $v=246$ GeV.

%
\section{Appendix:  Trick to reduce scalar-fermion elastic channel}
\label{app:trick}
%

In the elastic scatterings of species $a$ off some species $n$ in thermal equilibrium we encounter elastic rate function~\eqref{eq:gammaE}. The species in equilibrium follows the Maxwell-Boltzmann distribution:
\begin{equation}
    f^n_{\rm eq} = e^{-\beta E^n} \;,
\end{equation}
This allows us to perform the integration over the momentum, without needing to specify the functional form of the elastic cross section, using the following result:
\begin{align}
\Gamma^{an}_{\rm E}(p_1,T) &\equiv 
  \frac{1}{2\pi^2}\int_{0}^\infty {\rm d}p_3p_3^2 \;     
        f^n_{\rm eq}(p_3,T) \,[v_\M\sigma ]^{an}_{\rm E}(p_1,p_3) \;, \nonumber \\
    & = \frac{1}{16\pi^2p_1E_1} \int_{m_n}^{\infty} {\rm d}E_3
        e^{-\beta E_3}\;
        \int_{s_-}^{s_+} {\rm d}s \,
        \lambda^{1/2}(s,m_n^2,m_a^2) \, \sigma_{\rm E}^{an} (s) \;,
\label{eq:trick1}
\end{align}
where  $s_\pm =  m_n^2 + m_a^2 + 2E_1E_3 \pm 2 p_1p_3$. Then using identity $e^{-\beta E_3} = -T\frac{ \partial }{\partial E_3} e^{-\beta E_3}$, integrating by parts
and using the Leibniz integral rule and $\frac{{\rm d} s_{\pm}}{{\rm d}E_3} = 2(E_1 \pm E_3p_1/p_3)$, we find
\begin{align}
\Gamma^{an}_{\rm E}(p_1,T) 
    & = \frac{T}{8\pi^2p_1E_1} \int_{m_n}^{\infty} {\rm d}E_3
        e^{-\beta E_3}\;
        \left[
        \frac{{\rm d} s}{{\rm d}E_3} \;
        \lambda^{1/2}(s,m_n^2,m_a^2) \, \sigma_{\rm E}^{an} (s) 
        \right]_{s_-}^{s_+}  \nonumber \\
    & = \frac{T}{8\pi^2} \int_{0}^{\infty} {\rm d}p_3 \;
        e^{-\beta E_3}
        \left[
        \frac{p_3}{p_1E_3}F^-_{an}(p_1,p_3) + \frac{1}{E_1} F^+_{an}(p_1,p_3) \right] \;,
\label{eq:trick2}
\end{align}
where $F^\pm_{an}(p_1,p_3) = F_{an}(s_+) \pm F_{an}(s_-)$ and $F_{an}(s) \equiv \lambda^{1/2}(s,m_n^2,m_a^2) \, \sigma_{\rm E}^{an}(s)$. Note that the first term stays finite, as $s_+ \rightarrow s_-$ when $p_1 \rightarrow 0 $. The result~\eqref{eq:trick2} expresses
elastic scattering rate as a simple one-dimensional integral that can be computed and fitted before the integration of the Boltzmann equations.

%
\section{Appendix:  On the accuracy of the generalized relaxation time approximation}
\label{sec:appC}
%

At the core of our method is the generalized relaxation time approximation for the elastic collision integrals, introduced in section~\ref{sec:relax}. The accuracy of this scheme is not controlled by any small parameter, but similarly to the usual relaxation time approximation it should work well whenever a significant deviation from equilibrium is present, because the backward scattering terms are smoothed convolutions over the perturbation. One does not expect a high relative accuracy when deviation is small and/or smooth, but such deviations are irrelevant for the abundance calculation, because inelastic rates are then already accurately captured by a distribution with an equilibrium form.

We can verify these statements by a direct comparison to exactly computed elastic integrals. Instead of implementing simulation with full elastic integrals, we first do the calculation using our approximation scheme, saving the distribution function at each time-step. After this we evaluate the (kept) forward and the (dropped) backward elastic scattering terms numerically for the saved solutions. This allows us to evaluate the relative accuracy of our approach. We keep only the elastic tau channel elastic rate for this comparison.

%
\begin{figure}[t!]
\begin{center}
\includegraphics[width=0.306\textwidth]{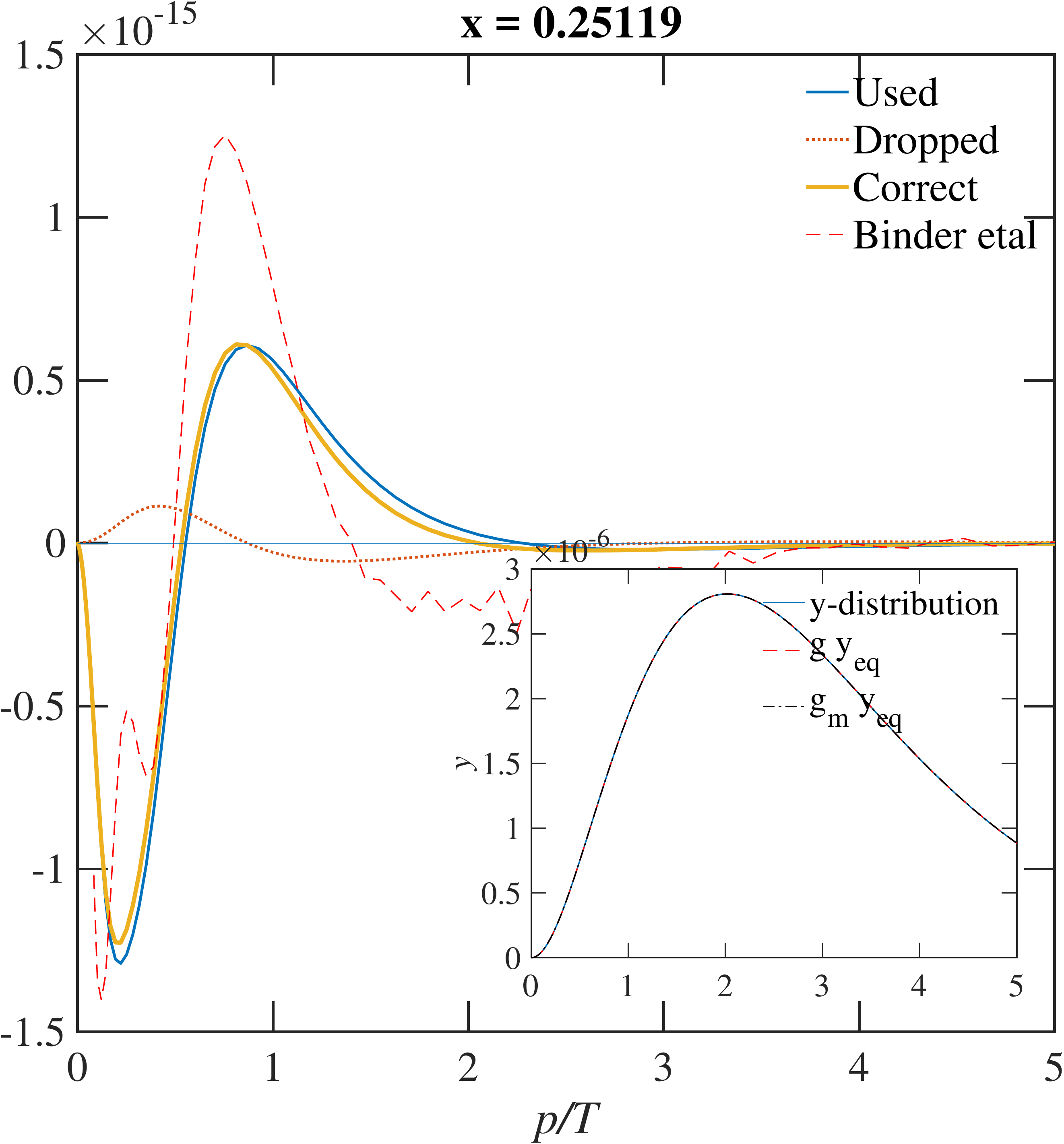} 
\includegraphics[width=0.30\textwidth]{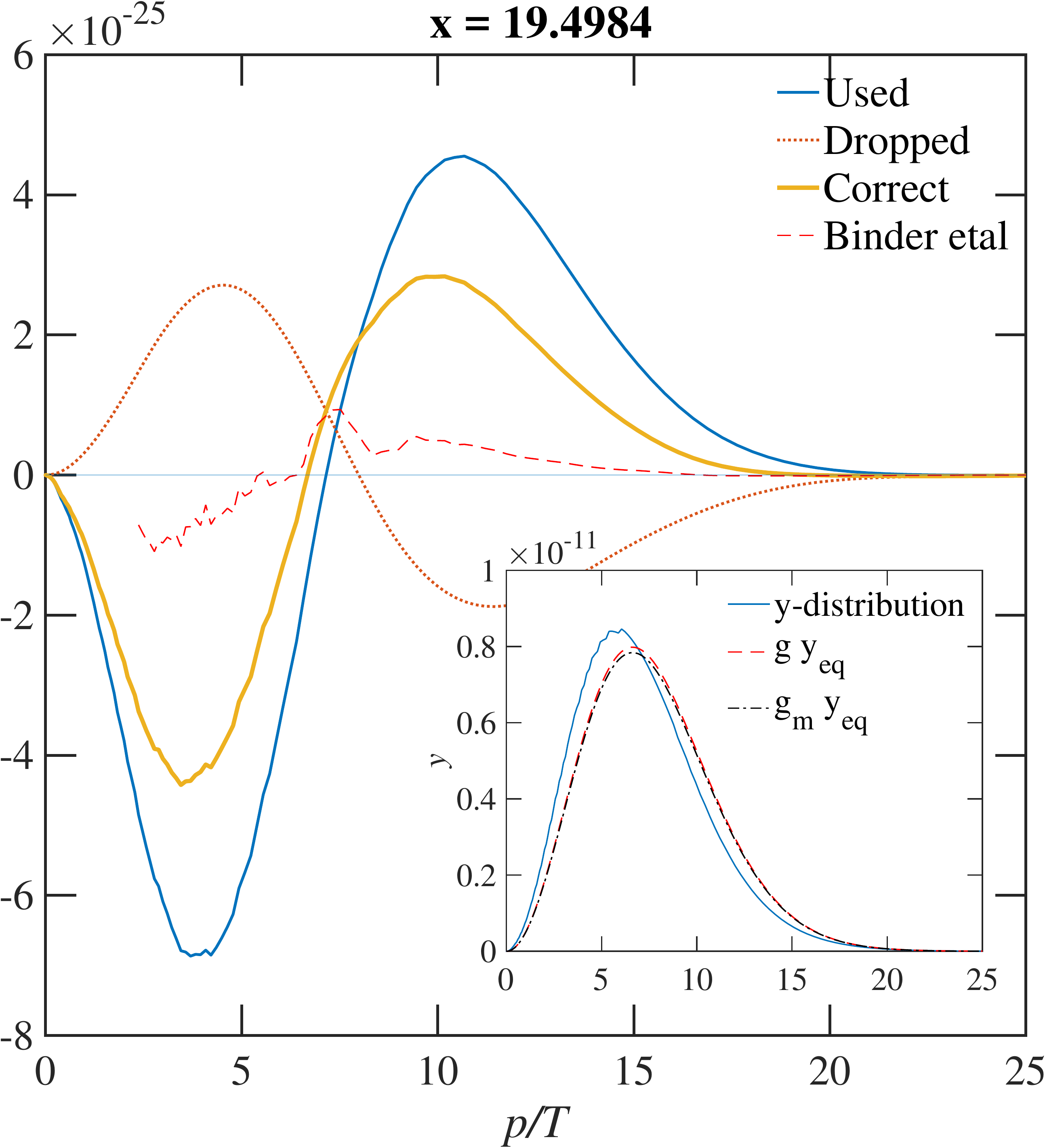}
\includegraphics[width=0.30\textwidth]{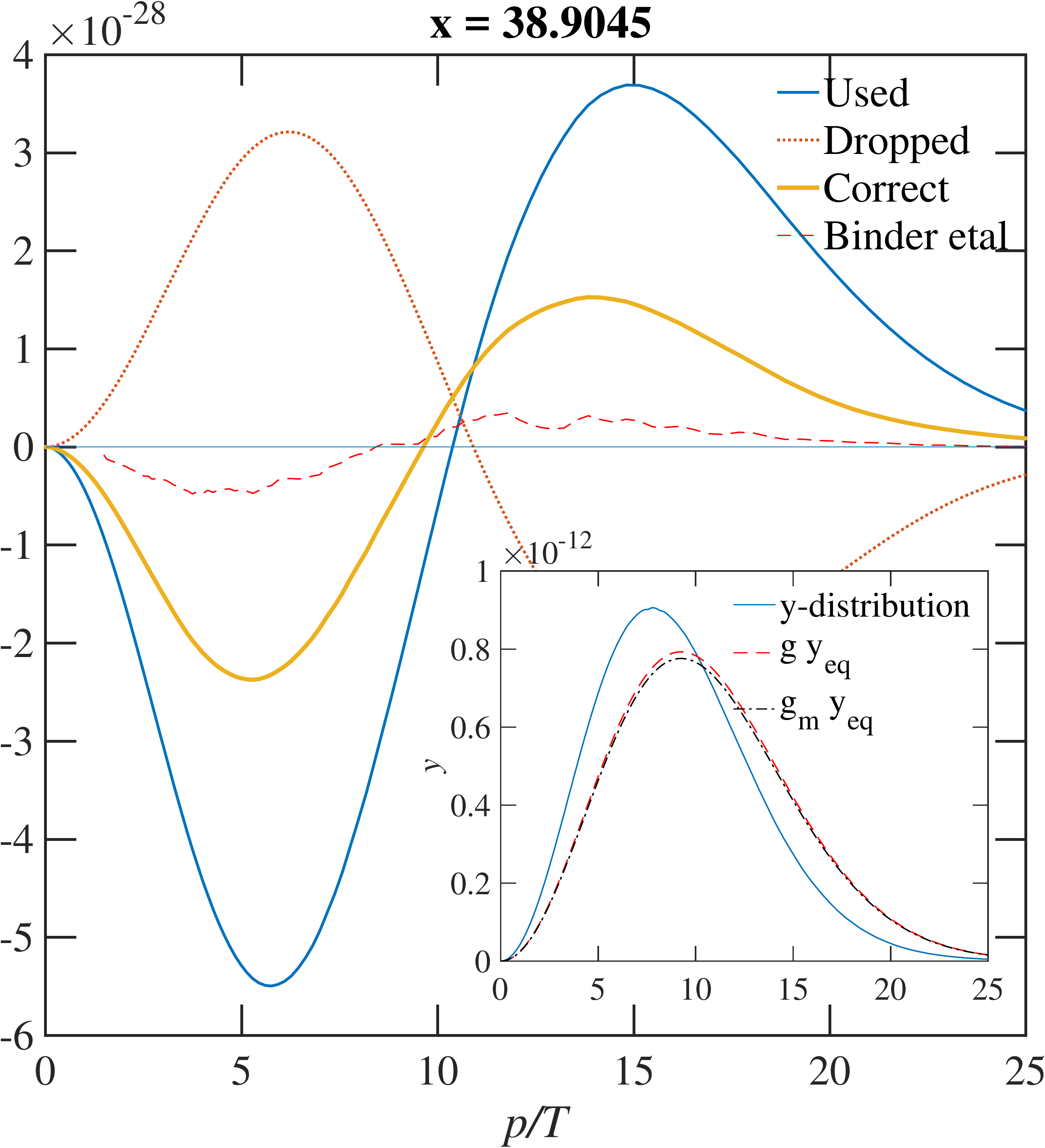}
\caption{Shown are (main frames) the elastic collision integrals $(p^2/2\pi^2)C_{\rm el}(p)$ as a function of the physical momentum for the distribution functions drawn from the sample calculation as indicated in figure~\ref{fig:figure9}. Each frame provides a snapshot at $x = m_s/T$ with $x$ given in the title. Blue lines correspond to the elastic integrals actually used in the calculation,  the dotted lines to the dropped back-scattering term and thick orange lines are the correct elastic integrals. Red dashed lines correspond to approximation~\eqref{eq:binder}. In the insets we show the actual distribution $y$ (solid blue line), as well as the scaled equilibrium distributions $gy_{\rm eq}$ (red dashed line) and $g_\tau y_{\rm eq}$ (black dash-dotted line).}
\label{fig:figure8}
\end{center}
\end{figure}
%

For a given distribution function $f(p,x)$ at the instant $x$, we first compute the $g_\tau(x)$-factor as defined in equation~\eqref{eq:gammam}, and then the deviation $\delta f = f - g_\tau f_{\rm eq}$. We then construct the forward and backward scattering terms exactly for this deviation. The former is of course given by equation~\eqref{eq:Celast}, while the latter one can be formally written as a convolution:
\be
\hat C^{\rm BW}_{\rm E}(\delta f; p_1,x) = \int{\rm d}p_3 G(p_1,p_3,x)  \delta f(p_3,x),
\label{eqlast1}
\ee
where the equilibrium function $G$ is defined as
\be
G(p_1,p_2,x) = \frac{p_3^2}{64\pi^5E_1}\int {\rm d}\Omega_3\int \frac{{\rm d}^3 p_2}{2E_2}\frac{{\rm d}^3 p_4}{2E_4} \delta^4(p_1+p_3-p_2-p_4)
|{\cal M}^{(\tau)}|^2 f_{\tau}(p_4,t).
\ee
where the matrix element for any $Sf\rightarrow Sf$ scattering is (for $\tau$ we have $N_\tau = 1$)
\be
|{\cal M}^{(f)}|^2 = 4N_f \lambda_{\rm hs}^2 m_f^2 (4m_f^2 - t)/(t - m_h^2)^2.
\label{eq:taumel}
\ee
We use the method introduced in ref.~\cite{Hannestad:1995rs} to reduce $G$ into a simple  integral over the magnitude of the 3-momentum $p_4$ and an additional angle (there are two angles if the matrix element depends also on the Mandelstam variable $s$). 

Having achieved this construction, we checked that the forward and the backward terms are properly normalised by comparing their magnitudes at full equilibrium, and that their integrals each vanish separately for the perturbation $\delta f$ as they should: 
\be
\int {\rm d}p_1p_1^2 \delta f(p_1,x) \Gamma_E(p_1,x) = \int {\rm d}p_1p_1^2 C_E^{\rm BW}(\delta f;p_1,x) 
= 0.\ee
This gives an additional check to all our formulae associated with the decay rates. As stated already, the only approximation in our approach, beyond using the Maxwell-Boltzmann equilibrium distributions, corresponds to our dropping the backward scattering term defined precisely in~\eqref{eqlast1}; for the scheme to work this term should be smaller than the forward term we used. We plot these terms in figure~\ref{fig:figure8} for a particular realization with the tau-channel only and with parameters $m_s=59$ GeV and $\lambda_{hs} = 0.0013$.

As is clear from figure~\ref{fig:figure8}, for small $x\approx 0.25$, where the out-of-equilibrium feature is sharp, our approximation is excellent, as expected. For $x \approx 19.5$ close to the freeze-out point, the errors are still reasonable, at most 30-40 percent. Even for a very large $x \approx 38.9$ and beyond, the approximation remains typically good to a factor of 2, although beyond the freeze-out this difference is irrelevant for the final result. Note that both the magnitude of the elastic integrals and that of the equilibrium distributions shown in the inset, vary over several orders of magnitude during the calculation ($g$ and $g_\tau$ are defined in equations~\eqref{eq:pseudoeqdist}  and~\eqref{eq:gammam} respectively).

\begin{figure}[t!]
\begin{center}
\includegraphics[width=0.7\textwidth]{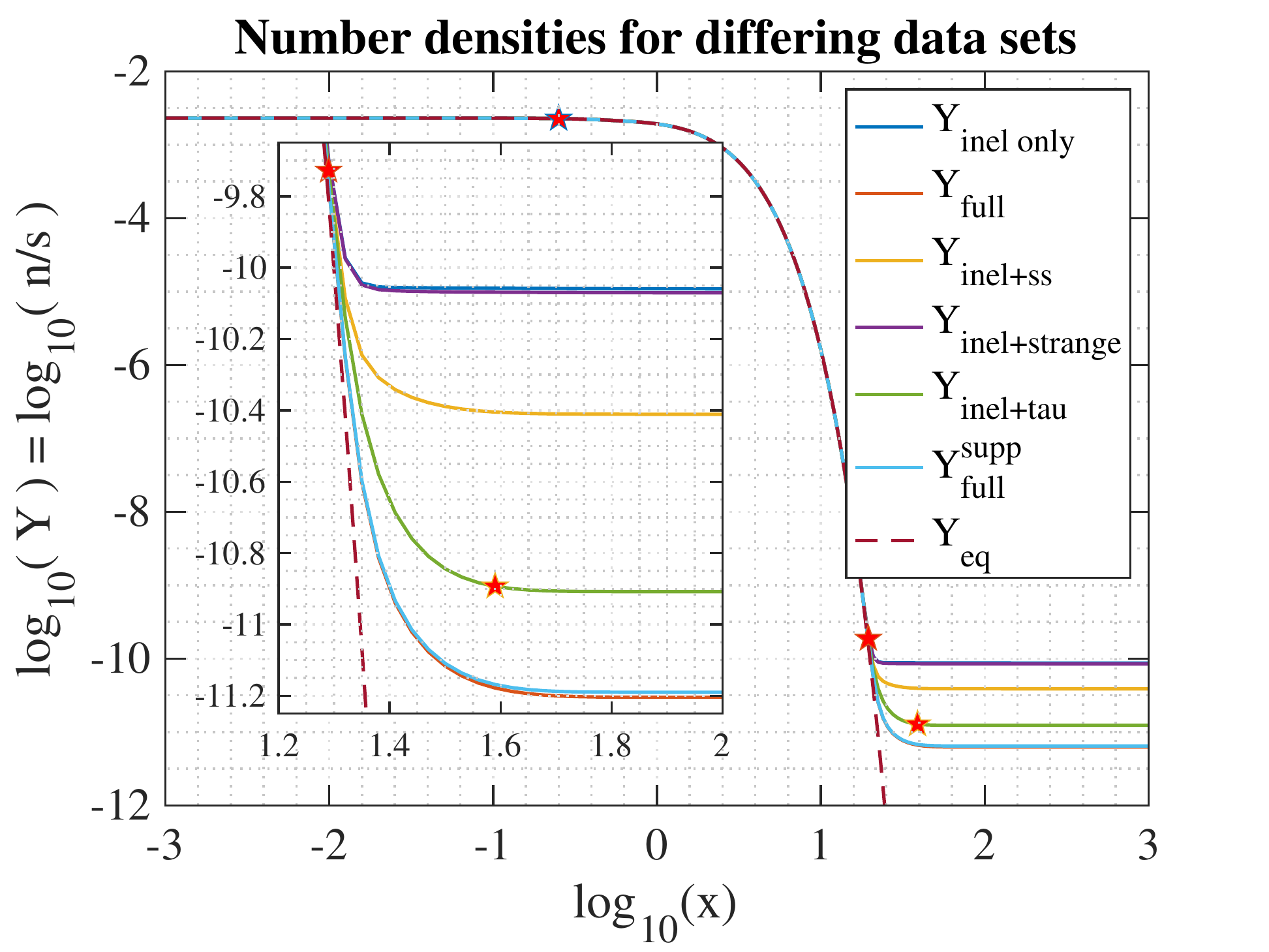} 
\caption{Singlet scalar yields calculated using various elastic channels in the computation. Reading the legend from up to down: the first five labels correspond to the scenarios visualised in figure~\ref{fig:BE2}, $Y_{\rm full}^{\rm supp}$ corresponds to the full computation where the elastic rates have been re-scaled downwards (see the text) and $Y_{\rm eq}$ visualises the equilibrium yield. Red markers denote the locations of the example points shown in figure~\ref{fig:figure9}. }
\label{fig:figure9}
\end{center}
\end{figure}

In general our method slightly over-estimates the elastic channel. To quantify the effect of this deviation we re-ran our code for our test case with elastic rates multiplied by 0.6. This increased the final abundance changed by 3 percent, which we believe is a conservative upper bound on the error.  In figure~\ref{fig:figure9} we visualise the yields $Y=n/s$  corresponding to scenarios detailed in figure~\ref{fig:BE2}, including the full computation with the re-scaled rates. Red markers in the figure~\ref{fig:figure9} show the points where we extracted the profiles in Figures~\ref{fig:figure8}.

%
%

Finally, we also compared our elastic integrals to the semirelativistic zero-momentum exchange approximation used in ref~\cite{Binder:2017rgn}. We reproduce their formula for the scattering rate here:
\be
C_{el} \approx \frac{\gamma(T)}{2}\left[ET \partial_p^2 + \left(p + 2T\frac{E}{p} + T\frac{p}{E} \right)\partial_p + 3\right]f(p,x),
\label{eq:binder}
\ee
where $E = \sqrt{p^2 + m_z^2}$. In this simple case
\be
\gamma(T) =\frac{m_s}{4} \int_{m_f}^\infty e^{-\omega/T}\Big(1-\frac{(m_s^2 - m_f^2)^2}{s^2}\Big)(-t_{\rm in})|{\cal M}(t_{\rm in})|^2,
\ee
where the matrix element $|{\cal M}(t)|^2$ is given in Eq.~\eqref{eq:taumel} (this differs by a factor 8 from Eq.~(42) in ref.~\cite{Binder:2017rgn}), $t_{\rm in} = (s-(m_s+m_f)^2)(s-(m_s-m_f)^2)/s$ and finally $s \approx m_s^2 + m_f^2 + 2m_s\omega$. This rate is shown by red dashed lines in figure~\ref{fig:figure8}. While our method slightly overestimates the elastic rate, the  approximation~\eqref{eq:binder}, typically underestimates it (the case displayed in the left panel is actually beyond the range of the validity of~\eqref{eq:binder}). The curves corresponding to~\eqref{eq:binder} also contain noise (even after some small-scale data-smoothing and/or using large step sizes) that comes from computing derivatives of a discrete distribution function. Regardless, based on our test runs, using~\eqref{eq:binder} is less accurate than our scheme.

We conclude that our method is a very good approximation for computing abundances to high precision. However, it should be applied with care to problems where a high-resolution final state momentum distribution is of prime importance. In such cases its accuracy should at least be tested by use of a exact momentum integrals.

%
\bibliography{Momentum.bib}

\providecommand{\href}[2]{#2}\begingroup\raggedright\begin{thebibliography}{10}

\bibitem{McDonald:1993ex}
J.~McDonald, \emph{{Gauge singlet scalars as cold dark matter}},
  \href{https://doi.org/10.1103/PhysRevD.50.3637}{\emph{Phys.Rev.} {\bfseries
  D50} (1994) 3637} [\href{https://arxiv.org/abs/hep-ph/0702143}{{\ttfamily
  hep-ph/0702143}}].

\bibitem{Cline:2013gha}
J.~M. Cline, K.~Kainulainen, P.~Scott and C.~Weniger, \emph{{Update on scalar
  singlet dark matter}},
  \href{https://doi.org/10.1103/PhysRevD.88.055025}{\emph{Phys.Rev.} {\bfseries
  D88} (2013) 055025} [\href{https://arxiv.org/abs/1306.4710}{{\ttfamily
  1306.4710}}].

\bibitem{Zeldovich:1965}
Y.~Zel'dovich, L.~Okun and S.~Pikelner, \emph{{Quarks: astrophysical and
  physicochemical aspects}}, {\emph{Sov. Phys. Uspekhi.} {\bfseries 8} (1966)
  702}.

\bibitem{Lee:1977ua}
B.~W. Lee and S.~Weinberg, \emph{{Cosmological Lower Bound on Heavy Neutrino
  Masses}},
  \href{https://doi.org/10.1103/PhysRevLett.39.165}{\emph{Phys.Rev.Lett.}
  {\bfseries 39} (1977) 165}.

\bibitem{Gondolo:1990dk}
P.~Gondolo and G.~Gelmini, \emph{{Cosmic abundances of stable particles:
  Improved analysis}},
  \href{https://doi.org/10.1016/0550-3213(91)90438-4}{\emph{Nucl. Phys.}
  {\bfseries B360} (1991) 145}.

\bibitem{Srednicki:1988ce}
M.~Srednicki, R.~Watkins and K.~A. Olive, \emph{{Calculations of Relic
  Densities in the Early Universe}},
  \href{https://doi.org/10.1016/0550-3213(88)90099-5}{\emph{Nucl. Phys.}
  {\bfseries B310} (1988) 693}.

\bibitem{Laine:2006cp}
M.~Laine and Y.~Schroder, \emph{{Quark mass thresholds in QCD thermodynamics}},
  \href{https://doi.org/10.1103/PhysRevD.73.085009}{\emph{Phys. Rev.}
  {\bfseries D73} (2006) 085009}
  [\href{https://arxiv.org/abs/hep-ph/0603048}{{\ttfamily hep-ph/0603048}}].

\bibitem{Drees:2015exa}
M.~Drees, F.~Hajkarim and E.~R. Schmitz, \emph{{The Effects of QCD Equation of
  State on the Relic Density of WIMP Dark Matter}},
  \href{https://doi.org/10.1088/1475-7516/2015/06/025}{\emph{JCAP} {\bfseries
  1506} (2015) 025} [\href{https://arxiv.org/abs/1503.03513}{{\ttfamily
  1503.03513}}].

\bibitem{Steigman:2012nb}
G.~Steigman, B.~Dasgupta and J.~F. Beacom, \emph{{Precise Relic WIMP Abundance
  and its Impact on Searches for Dark Matter Annihilation}},
  \href{https://doi.org/10.1103/PhysRevD.86.023506}{\emph{Phys.Rev.} {\bfseries
  D86} (2012) 023506} [\href{https://arxiv.org/abs/1204.3622}{{\ttfamily
  1204.3622}}].

\bibitem{Enqvist:1988we}
K.~Enqvist, K.~Kainulainen and J.~Maalampi, \emph{{Cosmic Abundances of Very
  Heavy Neutrinos}},
  \href{https://doi.org/10.1016/0550-3213(89)90537-3}{\emph{Nucl. Phys.}
  {\bfseries B317} (1989) 647}.

\bibitem{Cline:2012hg}
J.~M. Cline and K.~Kainulainen, \emph{{Electroweak baryogenesis and dark matter
  from a singlet Higgs}},
  \href{https://doi.org/10.1088/1475-7516/2013/01/012}{\emph{JCAP} {\bfseries
  1301} (2013) 012} [\href{https://arxiv.org/abs/1210.4196}{{\ttfamily
  1210.4196}}].

\bibitem{Dittmaier:2011ti}
{\scshape LHC Higgs Cross Section Working Group} collaboration, \emph{{Handbook
  of LHC Higgs Cross Sections: 1. Inclusive Observables}},
  \href{https://arxiv.org/abs/1101.0593}{{\ttfamily 1101.0593}}.

\bibitem{Planck:2015xua}
{\scshape Planck} collaboration, \emph{{Planck 2015 results. XIII. Cosmological
  parameters}},  \href{https://arxiv.org/abs/1502.01589}{{\ttfamily
  1502.01589}}.

\bibitem{Aprile:2018dbl}
{\scshape XENON} collaboration, \emph{{Dark Matter Search Results from a One
  Ton-Year Exposure of XENON1T}},
  \href{https://doi.org/10.1103/PhysRevLett.121.111302}{\emph{Phys. Rev. Lett.}
  {\bfseries 121} (2018) 111302}
  [\href{https://arxiv.org/abs/1805.12562}{{\ttfamily 1805.12562}}].

\bibitem{Griest:1990kh}
K.~Griest and D.~Seckel, \emph{{Three exceptions in the calculation of relic
  abundances}}, \href{https://doi.org/10.1103/PhysRevD.43.3191}{\emph{Phys.
  Rev.} {\bfseries D43} (1991) 3191}.

\bibitem{Binder:2017rgn}
T.~Binder, T.~Bringmann, M.~Gustafsson and A.~Hryczuk, \emph{{Early kinetic
  decoupling of dark matter: when the standard way of calculating the thermal
  relic density fails}},
  \href{https://doi.org/10.1103/PhysRevD.96.115010}{\emph{Phys. Rev.}
  {\bfseries D96} (2017) 115010}
  [\href{https://arxiv.org/abs/1706.07433}{{\ttfamily 1706.07433}}].

\bibitem{Bernal:2017kxu}
N.~Bernal, M.~Heikinheimo, T.~Tenkanen, K.~Tuominen and V.~Vaskonen, \emph{{The
  Dawn of FIMP Dark Matter: A Review of Models and Constraints}},
  \href{https://doi.org/10.1142/S0217751X1730023X}{\emph{Int. J. Mod. Phys.}
  {\bfseries A32} (2017) 1730023}
  [\href{https://arxiv.org/abs/1706.07442}{{\ttfamily 1706.07442}}].

\bibitem{Belanger:2018ccd}
G.~Bélanger, F.~Boudjema, A.~Goudelis, A.~Pukhov and B.~Zaldivar,
  \emph{{micrOMEGAs5.0 : Freeze-in}},
  \href{https://doi.org/10.1016/j.cpc.2018.04.027}{\emph{Comput. Phys. Commun.}
  {\bfseries 231} (2018) 173}
  [\href{https://arxiv.org/abs/1801.03509}{{\ttfamily 1801.03509}}].

\bibitem{Bernal:2018kcw}
N.~Bernal, C.~Cosme, T.~Tenkanen and V.~Vaskonen, \emph{{Scalar singlet dark
  matter in non-standard cosmologies}},
  \href{https://doi.org/10.1140/epjc/s10052-019-6550-9}{\emph{Eur. Phys. J.}
  {\bfseries C79} (2019) 30}
  [\href{https://arxiv.org/abs/1806.11122}{{\ttfamily 1806.11122}}].

\bibitem{Merle:2015oja}
A.~Merle and M.~Totzauer, \emph{{keV Sterile Neutrino Dark Matter from Singlet
  Scalar Decays: Basic Concepts and Subtle Features}},
  \href{https://doi.org/10.1088/1475-7516/2015/06/011}{\emph{JCAP} {\bfseries
  1506} (2015) 011} [\href{https://arxiv.org/abs/1502.01011}{{\ttfamily
  1502.01011}}].

\bibitem{Bernal:2015xba}
N.~Bernal and X.~Chu, \emph{{$\mathbb {Z}_2$ SIMP Dark Matter}},
  \href{https://doi.org/10.1088/1475-7516/2016/01/006}{\emph{JCAP} {\bfseries
  1601} (2016) 006} [\href{https://arxiv.org/abs/1510.08527}{{\ttfamily
  1510.08527}}].

\bibitem{Konig:2016dzg}
J.~K{\"o}nig, A.~Merle and M.~Totzauer, \emph{{keV Sterile Neutrino Dark Matter
  from Singlet Scalar Decays: The Most General Case}},
  \href{https://doi.org/10.1088/1475-7516/2016/11/038}{\emph{JCAP} {\bfseries
  1611} (2016) 038} [\href{https://arxiv.org/abs/1609.01289}{{\ttfamily
  1609.01289}}].

\bibitem{Heikinheimo:2017ofk}
M.~Heikinheimo, T.~Tenkanen and K.~Tuominen, \emph{{WIMP miracle of the second
  kind}}, \href{https://doi.org/10.1103/PhysRevD.96.023001}{\emph{Phys. Rev.}
  {\bfseries D96} (2017) 023001}
  [\href{https://arxiv.org/abs/1704.05359}{{\ttfamily 1704.05359}}].

\bibitem{Heikinheimo:2018esa}
M.~Heikinheimo, K.~Tuominen and K.~Langæble, \emph{{Hidden strongly
  interacting massive particles}},
  \href{https://doi.org/10.1103/PhysRevD.97.095040}{\emph{Phys. Rev.}
  {\bfseries D97} (2018) 095040}
  [\href{https://arxiv.org/abs/1803.07518}{{\ttfamily 1803.07518}}].

\bibitem{Heikinheimo:2016yds}
M.~Heikinheimo, T.~Tenkanen, K.~Tuominen and V.~Vaskonen, \emph{{Observational
  Constraints on Decoupled Hidden Sectors}},
  \href{https://doi.org/10.1103/PhysRevD.96.109902,
  10.1103/PhysRevD.94.063506}{\emph{Phys. Rev.} {\bfseries D94} (2016) 063506}
  [\href{https://arxiv.org/abs/1604.02401}{{\ttfamily 1604.02401}}].

\bibitem{Murgia:2017lwo}
R.~Murgia, A.~Merle, M.~Viel, M.~Totzauer and A.~Schneider, \emph{{"Non-cold"
  dark matter at small scales: a general approach}},
  \href{https://doi.org/10.1088/1475-7516/2017/11/046}{\emph{JCAP} {\bfseries
  1711} (2017) 046} [\href{https://arxiv.org/abs/1704.07838}{{\ttfamily
  1704.07838}}].

\bibitem{Murgia:2018now}
R.~Murgia, V.~Iršič and M.~Viel, \emph{{Novel constraints on noncold,
  nonthermal dark matter from Lyman-$\alpha$ forest data}},
  \href{https://doi.org/10.1103/PhysRevD.98.083540}{\emph{Phys. Rev.}
  {\bfseries D98} (2018) 083540}
  [\href{https://arxiv.org/abs/1806.08371}{{\ttfamily 1806.08371}}].

\bibitem{Hannestad:1995rs}
S.~Hannestad and J.~Madsen, \emph{{Neutrino decoupling in the early universe}},
  \href{https://doi.org/10.1103/PhysRevD.52.1764}{\emph{Phys. Rev.} {\bfseries
  D52} (1995) 1764} [\href{https://arxiv.org/abs/astro-ph/9506015}{{\ttfamily
  astro-ph/9506015}}].

\end{thebibliography}\endgroup



\providecommand{\href}[2]{#2}\begingroup\raggedright\endgroup
%

%
\end{document}